\documentclass[preprint,12pt]{elsarticle}

\usepackage{graphicx,amsmath,amssymb,url}
\usepackage[usenames]{color}
\usepackage[citecolor=blue]{hyperref}
\hypersetup{
    colorlinks=true,
    linkcolor=blue,
    urlcolor=cyan,
}

\journal{Theoretical Computer Science}

\newtheorem{proposition}{Proposition}[section]
\newtheorem{theorem}[proposition]{Theorem}
\newtheorem{corollary}[proposition]{Corollary}
\newtheorem{lemma}[proposition]{Lemma}
\newtheorem{remark}[proposition]{Remark}

\newproof{pf}{Proof}

\newcommand{\wt}[1]{\widetilde{#1}}
\newcommand{\dt}[1]{{\bf #1}}
\newcommand{\ignore}[1]{}	

\newcommand{\mt}{\rightarrow}

\newcommand{\ol}[1]{\overline{#1}}

\newcommand{\beq}{\begin{equation}}
\newcommand{\eeq}{\end{equation}}
\newcommand{\beql}[1]{\begin{equation}\label{eq:#1}}
\newcommand{\eeql}{\end{equation}}
\newcommand{\beqarrays}{\begin{eqnarray*}}
\newcommand{\eeqarrays}{\end{eqnarray*}}

\newcommand{\blem}{\begin{lemma}}
\newcommand{\elem}{\end{lemma}}
\newcommand{\bleml}[1]{\begin{lemma} \label{lem:#1}}
\newcommand{\eleml}{\end{lemma}}
\newcommand{\blemT}[2]{\begin{lemma}[#1] \label{lem:#2}}
\newcommand{\elemT}{\end{lemma}}
\newcommand{\bthm}{\begin{theorem}}
\newcommand{\ethm}{\end{theorem}}
\newcommand{\bthml}[1]{\begin{theorem} \label{thm:#1}}
\newcommand{\ethml}{\end{theorem}}
\newcommand{\bthmT}[2]{\begin{theorem}[#1] \label{thm:#2}}
\newcommand{\ethmT}{\end{theorem}}
\newcommand{\brem}{\begin{remark}}
\newcommand{\erem}{\end{remark}}
\newcommand{\breml}[1]{\begin{remark} \label{rem:#1}}
\newcommand{\ereml}{\end{remark}}
\newcommand{\bcor}{\begin{corollary}}
\newcommand{\ecor}{\end{corollary}}
\newcommand{\bcorl}[1]{\begin{corollary} \label{cor:#1}}
\newcommand{\ecorl}{\end{corollary}}
\newcommand{\bpropl}[1]{\begin{proposition} \label{pro:#1}}
\newcommand{\epropl}{\end{proposition}}

\newcommand{\bpf}{\begin{pf}}
\newcommand{\epf}{\end{pf}\hfill \qed}

\newcommand{\refeq}[1]{(\protect\ref{eq:#1})}

\newcommand{\refLem}[1]{Lemma~\protect\ref{lem:#1}}
\newcommand{\refThm}[1]{Theorem~\protect\ref{thm:#1}}
\newcommand{\refFig}[1]{Figure~\protect\ref{fig:#1}}
\newcommand{\refCor}[1]{Corollary~\protect\ref{cor:#1}}

\newcommand{\refSec}[1]{Section~\protect\ref{sec:#1}}

\newcommand{\refPro}[1]{Proposition~\protect\ref{pro:#1}}
\newcommand{\CC}{{\mathbb C}}
\newcommand{\NN}{{\mathbb N}}

\newcommand{\RR}{{\mathbb R}}
\newcommand{\ZZ}{{\mathbb Z}}
\newcommand{\FF}{{\mathbb F}}
\newcommand{\as}{\textcolor{red}{\mathop{\mbox{\rm :=}}}}
\newcommand{\ras}{\textcolor{red}{\mathrel{\,=:\,}}} 
\newcommand{\dd}{ ,\ldots , }

\newcommand{\ib}{\subseteq }
\newcommand{\ibp}{\subset }		

\newcommand{\abs}[1]{\left\lvert#1\right\rvert}       
\newcommand{\es}{\emptyset}

\newcommand{\da}{\;\text{dA}}

\newcommand{\calA}{\mathcal{A}}
\newcommand{\calB}{\mathcal{B}}
\newcommand{\calc}{\mathcal{C}}
\newcommand{\calD}{\mathcal{D}}

\newcommand{\wtC}{\wt{C}}

\newcommand{\omin}{\ol{\min}}
\newcommand{\intbox}{{\,\,\setlength{\unitlength}{.33mm}\framebox(4,7){}\,}}
\newcommand{\set}[1]{\left\{ #1 \right\}}
\newcommand{\eps}{\epsilon}
\newcommand{\paren}[1]{\left( { #1 }\right)}	
\newcommand{\su}{\cup}
\newcommand{\si}{\cap}
\newcommand{\sm}{\setminus}
\newcommand{\Mea}{\mathcal{M}}
\newcommand{\ass}{\leftarrow}		
\newcommand{\nin}{\not\in}
\newcommand{\half}{\frac{1}{2}}	

	\newenvironment{prog}{\begin{tabbing}
	 xxxx\=xxxx\=xxxx\=xxxx\=xxxx\=xxxx\=xxxx\=xxxx\=xxxx\=xxxx\=xxxx\=xxxx\=xxxx\=
	\kill\\}{
		\end{tabbing}}

	\newenvironment{progb}[2][2]{ 
		\begin{center}
		  \fbox{\begin{minipage}{0.75\textwidth}
			\begin{prog}#2\end{prog}
	          \end{minipage}}
	        \end{center}
		}{}
	
	\newcommand{\cocyan}[1]{\textcolor{cyan}{#1}}	 
        \newcommand{\Comment}[1]{
          \quad\cocyan{\mbox{$\triangleleft\;\;$}{\em #1}} }

	\newcommand{\vfigpdf}[3]{
            \begin{figure}[htb]
              \centering
              \includegraphics[scale=#3]{#2}
              \caption{#1}
              \label{fig:#2}
            \end{figure}
          }

\begin{document}
\begin{frontmatter}
  \title{Complexity  of a Root Clustering Algorithm}
  \author{Prashant Batra}
  \address{Hamburg University of Technology, Germany.}
  \ead{batra@tuhh.de}
  \author{Vikram Sharma \corref{cor1}}
  \address{The Institute of Mathematical Sciences, HBNI, Chennai, India.}
  \ead{vikram@ims.res.in}
  \ead[url]{http://www.imsc.res.in/\~{} vikram}

  \begin{abstract}
    Approximating the roots of a holomorphic function in an input box is a fundamental problem in many domains.
    Most algorithms in the literature for solving this problem are conditional, i.e., 
    they make some simplifying assumptions, such as, all the roots are simple or there are no roots 
    on the boundary of the input box, or the underlying machine model is Real RAM.
    Root clustering is a generalization of the root approximation problem that allows for errors in the computation
    and makes no assumption on the multiplicity of the roots.
    An unconditional algorithm for computing a root clustering of a holomorphic function was given by
    Yap, Sagraloff and Sharma in 2013 \cite{yap-sagraloff-sharma:cluster:13}. They proposed a subdivision based algorithm using effective
    predicates based on Pellet's test while avoiding any comparison with zeros (using soft zero comparisons instead).
    In this paper, we analyze the running time of their algorithm. We use the continuous amortization
    framework to derive an upper bound on the size of the subdivision tree. We specialize this bound
    to the case of polynomials and some simple transcendental functions such as exponential and trigonometric
    sine. We show that the algorithm takes exponential time even for these
    simple functions, unlike the case of polynomials. We also derive a bound on the bit-precision used by the algorithm. To the best 
    of our knowledge, this is the  first such result for holomorphic functions. We introduce new geometric
    parameters, such as the relative growth of the function on the input box, 
    for analyzing the algorithm. Thus, our estimates naturally generalize the known results, i.e., for the case of polynomials.
  \end{abstract}
\end{frontmatter}


\section{Introduction}
\label{sec:introduction}
Numerical analysis is concerned with developing algorithms that work over a continuous domain,
such as the reals $\RR$ or the complex numbers $\CC$.
A fundamental problem in this area is
to approximate the roots of a 
holomorphic function $f: \CC \mt \CC$ in an input box $B_0$, with dyadic endpoints,
to within some desired input accuracy.
We call this problem the \dt{exact root finding} (ERF) problem.
We assume throughout that $f$ is holomorphic not only on $B_0$
but over a sufficiently large neighborhood of  $B_0$. 
We further assume that  $f$ and {\em all its  derivatives} are represented by 
their ``box''-versions $\intbox f^{(j)}$, $j \ge 0$,
i.e., given a box $B \ibp \CC$, we can compute a set $\intbox f^{(j)}(B)$ that includes the range $f^{(j)}(B)$;
moreover, for a sequence of boxes monotonically converging to a point the error in the
overestimation converges linearly to zero; such box functions are known for a large class of functions, e.g., \cite{du-yap:hyper:06}.
Alternatively, one can construct box-functions  using techniques such as automatic differentiation, 
given an algorithm for evaluating a box-function for the function \cite{griewank-walther:bk}.

The ERF  problem  naturally generalizes the problem of
approximating  the roots of a polynomial \cite{henrici:applied-bk,schonhage:fundamental}.
However,  there are very few ``complete'' algorithms
in the literature for solving it \cite{giusti+2:zeros-analytic:05,johnson-tucker:analytic-zeros:09}.
The few that are algorithmically sound, do not have a rigorous complexity analysis.
The occasional results implying a complexity analysis are not complete:
One such result states that the roots of a polynomial-time 
computable function are also poly-time computable \cite[Thm. 4.11]{ko:bk};
this result, however, is not uniform 
and it assumes that all the roots are  simple. Most of the algorithms in the literature make
similar assumptions: e.g., algorithms relying on the argument principle to detect roots in a domain
often assume that there is no root on the boundary of the domain \cite{johnson-tucker:analytic-zeros:09}; or algorithms that rely on
subdivision, either of the search domain  \cite{yakoubsohn:bisection-analysis:05}
or while applying quadrature rules \cite{ying-katz:zeros-analyic:88}, often
introduce some preset tolerance to stop the subdivision.
The situation is in contrast to the setting of polynomials where many complete algorithms are
known with detailed and near optimal complexity estimates \cite{schonhage:fundamental,pan-near-optimal:2002,becker+3:cisolate:18}.

A desirable goal, therefore, is to devise a \dt{complete algorithm} for the ERF problem,
i.e., an algorithm that does not make such niceness assumptions. 
However, working with a box-representation of $f$ means that we cannot distinguish,
using finite precision,
whether a region contains a multiple root or  a ``cluster'' of very close roots.
In order to address this we have to change the problem from approximating roots to approximating ``clusters''. 
This is called the \dt{root clustering problem} (RCP) \cite{yap-sagraloff-sharma:cluster:13}.
A further restriction on the algorithm is imposed by the Turing machine model of computational:
since we can only compute absolute approximations to numbers,
it is undecidable to detect if a number is zero or not.
Hence, we cannot use  comparisons with zero in devising complete algorithms.
We instead use a \dt{soft zero test}, i.e.,  
a comparison that only decides the sign of a {\em non-zero real number},
similar to what is implemented in the \texttt{iRRAM} software package \cite{irram}. 
A complete algorithm based on soft zero tests for the RCP was presented  
in \cite{yap-sagraloff-sharma:cluster:13}. 
In this paper, we bound its complexity, including its precision requirements,
and consequently obtain a generalization of similar results  for the case of polynomials  \cite{becker+4:cluster:16}. 

To describe the problem, we need the following definitions. 
For a disc $D \ibp \CC$ and non-negative number  $\lambda$,
by $\lambda D$ we denote the centrally scaled version of $D$ by the scaling factor $\lambda$.
Similarly, for an axis-aligned box $B \ibp \CC$, and $\lambda > 0$, let $\lambda B$ denote 
its centrally scaled version. Let $m(B)$ denote the midpoint of $B$, $w(B)$ its width, and 
$r_B \as w(B)/\sqrt{2}$
the radius of the smallest disc, $D(B)$, that is centered at $m(B)$ and contains $B$.

A non-empty {\em multiset} $\calc$ of a subset of roots of $f$ is called a \dt{cluster} if there exists a disc 
~$D$ and a  $\lambda >1$ such that $\calc$ is the largest multiset of roots in $D$ and $\lambda D$. 
The cardinality, $|\calc|$, of $\calc$ is the
number of roots in $\calc$ counted with multiplicity; roots in a cluster will always be counted with
multiplicity. A cluster $\calc$ is called a \dt{natural cluster}
if $\calc$ is the largest multiset of roots  in $D$ and $3D$;
such a disc  $D$ is called \dt{isolating} for $f$.
A crucial observation is that  the set of natural clusters form
a tree under subset inclusion  \cite{yap-sagraloff-sharma:cluster:13}.
A disc $D$ is called an \dt{exclusion disc} if it does not contain any 
root of $f$. 
The \dt{local Root Clustering Problem} for $f$ in a box $B_0$ is to construct
a collection $\calD = \set{D_1 \dd D_n}$ of pairwise disjoint discs such that
\begin{enumerate}
\item the centers $m(D_i)$ lie in  $B_0$,
\item each $D_i$ is an isolating disc for $f$, 
\item all roots of $f$ in $B_0$ are in some disc in $\calD$, and
\item all the roots of $f$ in the discs $D_i$ are in  $2B_0$.
\end{enumerate}
One may additionally require that  the isolating discs have radius smaller than a given $\eps$, 
but in our case we do not restrict the discs (i.e., will assume $\eps$ to be infinity).

The algorithm in \cite{yap-sagraloff-sharma:cluster:13} takes as
input a box $B_0$ and an upper 
bound $N_0$ on the number of roots of $f$ in $B_0$ counted with multiplicities.
We assume throughout that $2B_0$ is contained within the domain of holomorphy of $f$.
The algorithm is  subdivision based, that is, it constructs a quad-tree
subdivision of the search space in order to localize the roots
(this strategy was also used in \cite{dellnitz+2:zero-subdivision:01,johnson-tucker:analytic-zeros:09,yakoubsohn:bisection-analysis:05}).
Given a box $B$, the algorithm first tries to count the number of roots in a scaled version of the box $B$ 
using Pellet's test: if the function $f$ satisfies the following condition
        \beq
        \abs{\frac{f^{(k)}(m)}{k!}} r^k \ge \sum_{j \ge 0; j \neq k} \abs{\frac{f^{(j)}(m)}{j!}} r^j,
        \eeq
then the disc $D(m,r)$ contains exactly $k$ roots of $f$. To make this test effective,
the algorithm uses an interval arithmetic version of a result of Darboux \cite{darboux:sur:87}
combined with a soft zero comparison.
The test either returns the exact
number of roots $k \ge 0$ inside the scaled version of $B$, 
or it returns saying that it cannot determine the number of roots in $B$;
note that $B$ itself may not contain the $k$ roots since the test is applied to a scaling of $B$. 
In the former scenario, we terminate the subdivision at $B$, and in the latter scenario we subdivide
$B$ into four equal boxes and continue recursively. 
The output of the algorithm is an isolating system $\calD$, where each disc
in $\calD$ corresponds to a pair $(B,k)$, and denotes
the smallest disc centered at $m(B)$ and containing $c_1k \cdot B$, for some constant $c_1 > 1$. 
The complete details of the algorithm can be found in \refSec{soft-zero-algo}.

Our two main results are a bound on the number of subdivisions executed by the algorithm,
and a bound on the precision requirements of the algorithm. These two, along with standard bounds
for polynomial arithmetic, yield a complete complexity analysis of the algorithm.
We use the general framework of  continuous amortization \cite{burr:contAmortization:16} to derive the
bound on the size of the subdivision tree. In order to state the bound, we introduce
some notation.

For $k \ge 0$, and $z \in \CC$, we use the notation $f_k(z)$ to stand for $f^{(k)}(z)/k!$.
Crucial to our analysis is Smale's $\gamma$-function \cite{bcss:bk}:
        \beql{gammaf}
        \gamma(f,z) \as \sup_{k \ge 1} \abs{\frac{f_k(z)}{f(z)}}^{1/k}.
        \eeql
For a cluster of roots $\calc$, we can factor $f(z) = h(z) \prod_{\alpha \in \calc}(z-\alpha)$, 
where $h$ is a holomorphic function with no roots in $\calc$. 
The $\gamma$-function associated with $\calc$ is defined as
        \beql{gammac}
        \gamma_{\calc}(z)\as \gamma(h,z).
        \eeql
Let 
        \beql{mcalc}
        m_\calc \as \text{ centroid of the points in $\calc$}
        \eeql
 and 
        \beql{rcalc}
        r_\calc \as \min\set{r: \text{ $D(m_\calc,r)$ contains all the points in $\calc$}}.
        \eeql
Define
        \beql{Rcalc}
        R_\calc \as \frac{|\calc|}{c_0\gamma_\calc(m_\calc)}
        \eeql
where $c_0 \as 2^{17}e^2$. Intuitively speaking, $R_\calc$ measures the distance from the centroid $m_\calc$
to the nearest root of the $|\calc|$-th derivative of $f$. A cluster $\calc$ is said to be
\dt{strongly separated} if 
        \beql{ssc}
        r_\calc \le \frac{R_\calc}{8|\calc|^3}.
        \eeql
All roots of the function, irrespective of their multiplicities, are strongly separated clusters
since $r_\calc =0$ for them. For a strongly separated cluster $\calc$ define the disc
        \beql{dc}
        D_\calc \as D\paren{m_\calc, \frac{R_\calc}{|\calc|^3}}. 
        \eeql
The size of the subdivision tree is not only governed by the roots in $B_0$ but also those in a larger neighborhood of
it. This is captured by the following definition: Given $B_0$, consider the following set of clusters
        \beql{s0}
        S_0 \as \set{
           \text{maximal strongly separated clusters $\calc$ partitioning the roots of $f$ in $2B_0$
             }}.
        \eeql
Here maximal is in terms of subset inclusion, i.e., there is no larger strongly separated cluster
satisfying the same conditions; $S_0$ captures those clusters that are in the vicinity of
the box $B_0$, though all of them may not be detected by the algorithm.\footnote{The proofs later
show that $2B_0$ can be replaced by $(1+\eps)B_0$, for a constant $\eps > 0$.}

The precise statement of our first main result is the following:
\bthml{main}
The size of the subdivision tree of the algorithm in \cite{yap-sagraloff-sharma:cluster:13} is
        $$O\paren{|S_0| + \int_{B_0 \sm \su_{\calc \in S_0}D_\calc/3} \gamma_f(z)^2 \da},$$
where $\da$ denotes the area form.
\ethml
As a corollary, we obtain that the size of the subdivision tree for the local root cluster problem
{\em in case $f$ is a polynomial} of degree $d$ and $B_0$ is the box of width $2\Mea(f)$ containing
all the roots of $f$ is bounded by $O(d^2 \log \Mea(f)+d^3)$, where
$\Mea(f)$ denotes the Mahler measure of $f$ \cite{yap:bk}.
This is better than the bound in \cite{yakoubsohn:bisection-analysis:05} in two respects: 
first, we save a factor in the degree, and second, we do not assume that the polynomial is square-free;
the improvement is possible because the continuous amortization framework
allows us to do an amortized application of root bounds. 
If $f$ is the exponential function, 
then $S_0$ is empty and the $\gamma$-function is a constant, which implies that the
number of boxes in the subdivision tree is proportional to the area of $B_0$.
This does not appear to be bad at first sight, but observe that the area of $B_0$ is exponential in 
the bit-representation of its vertices, and so the bound is exponential in the input.  
This bound is also tight, because for the exclusion test based on Pellet's condition to detect that 
a box has no roots, we require that  $\exp(m(B))$ dominates $r_B \cdot \exp(z)$, where $z$ is on
the boundary of $B$; for this to hold, the width of the box $B$ must be smaller than one;
this means any box with width larger than one has to be subdivided into a uniform grid of boxes
of width smaller than one before they can be excluded.
Similar results can be shown for the  sine function.
Our results highlight that exclusion tests which may work very well for polynomials do not 
necessarily yield good results in the general setting.

To express the bound on precision, we need the following definitions:
Let 
        $$M(f, 4B_0) \as \sup \set{|f(z)|: z \in D(4B_0)}$$
and 
        $$m(f, B_0) \as \inf \set{|f(z)|: z \in B_0 \sm \su_{\calc \in S_0}D_\calc/3}.$$
For a  set $U \ibp \RR$, define
        \beql{omin}
        \omin(U) \as \min\set{1, \inf U}.
        \eeql
If $U=\set{x}$ then we simply write $\omin(x)$; observe that for all $x \ge 0$, we have
$\log \omin(x) \le 0$. {\em We also assume that the box-functions
satisfy a first-order error bound} (see \refeq{convergence} below for a precise definition).
Then the bound on the  precision requirement is given by the following:
\bthml{precision}
The precision required by the procedure is bounded by 
\begin{align*}
          O\Big(N_0\log \frac{M(f, 4B_0)}{m(f, B_0)} + N_0^2 - N_0^2 \log \omin\paren{w(B_0)}
                -N_0(\log \omin_{{\calc \in S_0}} \abs{f_{|\calc|}(m_\calc)})\Big).
\end{align*}
Recall that $N_0$ is an upper bound on the number of roots of
$f$ in $B_0$ counted with multiplicities, and $w(B_0)$ denotes the width of $B_0$.
\ethml
The ratio in the first term in the bound above measures the relative growth of the function. This is the first time,
to the best of our knowledge, that such a term appears in the analysis of a continuous algorithm.
The term $(-\log \abs{f_{|\calc|}(m_\calc)})$, for a cluster $\calc$,
in the last term of the bound measures the conditioning of the cluster: if it is very large, then we expect
to have a root near the cluster. When  $f$ is a polynomial of degree
$d$, substituting $N_0= d$, we obtain a bound similar to the one given in \cite[Thm.~A]{becker+4:cluster:16}
(see \refCor{precisionpoly}). Similar parameters have also
been used for bounding bit-complexity of numerical operators on analytic functions 
in real analysis \cite{kawamura+3:bit-complexity:15}.

\subsection{Our Contributions and Comparison to Existing Literature}
\label{sec:comp-exist-liter}

Yakoubsohn's exclusion-based algorithm \cite{yakoubsohn:bisection-analysis:05} is a well-known algorithm for 
zero-finding of analytic functions and is the closest algorithm to the one proposed in \cite{yap-sagraloff-sharma:cluster:13}.
It uses exclusively an exclusion predicate based on Pellet's test, and does not have an inclusion test;
the termination is guaranteed by using a cutoff precision for the width of the boxes.
We hope that it will be  instructive to compare the two approaches and especially the analysis of the algorithms.

Both approaches need evaluation of an analytic function $f$ given by a Taylor expansion. 
Whereas the algorithm in \cite{yap-sagraloff-sharma:cluster:13} assumes the existence of box-functions for $f$ and its higher
order derivatives, it is not clear how  Pellet's test is made effective in \cite{yakoubsohn:bisection-analysis:05}
for non-polynomial functions in general.

	

Besides the box-functions, the other algorithmic requirement of \cite{yap-sagraloff-sharma:cluster:13} 
is an externally given upper bound $N_0$ on the number of zeros (counted with multiplicity) 
in the input box $B_0$. In this box roots will be clustered naturally, i.e.,
the algorithm determines isolating discs for natural clusters (as defined in the introduction). 
Yakoubsohn \cite{yakoubsohn:bisection-analysis:05} instead requires four conditions
on the initial square $B_0$ with width $w(B_0)$.
Let the zeros of $f$ inside $B_0$ be $z_1, \ldots, z_d$ (where
multiple roots occur multiple times according to their
multiplicity). With $g(z):= \prod_{k=1}^d(z-z_k)$ we define a
splitting of $f$, $f(z)=g(z)h(z)$, into analytic functions. The first
requirement is (condition (1) in
\cite{yakoubsohn:bisection-analysis:05}) 
        \beql{Yak-Con-1} 
        \forall x  \in B_0: \left | \frac{h^{(k)}(x)}{k!h(x)} \right | \leq \lambda \tau^{k-1} \quad \mbox{for all } \, k \in \mathbb{N}, 
        \eeql 
for some real values $\lambda$ and $\tau$. Second requirement (condition (2) in
\cite{yakoubsohn:bisection-analysis:05}) is 
        \beql{Yak-Con-2} 
        4\sqrt{2} \tau w(B_0) \leq 1.  
        \eeql 
For the precise statement of Yakoubsohn's remaining conditions (3) and (4) please see  
the original paper \cite{yakoubsohn:bisection-analysis:05};
in particular, condition (4) implies that $\tau$ must be positive.
While condition (3) encapsulates an inclusion/isolation ratio around a zero, 
the additional condition (4)  requires a certain ratio size in condition (3), and compares and adjusts this 
globally for the cluster size in relation to the isolation distance. 
These additional requirements might necessitate a regrouping of all
zeros in the analysis,  which may not be reflected in the output. A
finite precision analysis would have to show how any type of grouping
could remain stable under the algorithm in
\cite{yakoubsohn:bisection-analysis:05}.

Coming to the complexity analysis of the algorithm, the main differences between
\cite{yakoubsohn:bisection-analysis:05} and our results are the following:
\begin{itemize}
\item Unlike \cite{yakoubsohn:bisection-analysis:05}, where a bit-complexity result 
is given for the case of square-free polynomials only, we 
provide a complexity analysis for non-polynomial functions also. Even in the case of polynomials,
we do not assume square-freeness.

\item The analysis of the polynomial case in \cite{yakoubsohn:bisection-analysis:05}, i.e., $f \in \mathbb{C}[z], $ is achieved in \cite{yakoubsohn:bisection-analysis:05}  by setting the
parameter $\lambda$  to zero in the above requirements
(\emph{viz.} \cite{yakoubsohn:bisection-analysis:05}, Section 9). But
this amounts to $h \equiv cons.$, i.e., the clustering of all
roots. Especially, the case of a box $B_0$ containing only a proper
subset of roots cannot be analyzed in this fashion.
A simple
transition from $B_0$ to a box containing all roots of $f \in \mathbb{C}[z]$ 
seems out of reach as the used predicate is sensitive
to roots outside the box. Our result in the polynomial setting, \refThm{polybd}, does not need the
assumption that $B_0$ includes all the roots of the polynomial.

\item The complexity estimate in \cite{yakoubsohn:bisection-analysis:05} depends on
the distance of $f$ to the set of polynomials with multiple roots, unlike our results
which depend on natural parameters associated with the polynomial, such as the degree and
the Mahler measure.

\item The estimate of Theorem 1.1 in \cite{yakoubsohn:bisection-analysis:05}
can be specialized (see Section 9.1 of the paper) to yield a number of subdivisions
of order $O(d^4 \log \Mea(f) + d^4 \log d)$ for a polynomial of degree $d$. This is at
least a factor larger in the degree than our bound, \refCor{intpoly}; for polynomials
where $\Mea(f)$ dominates, the improvement is quadratic.

\item Unlike our results on the number of boxes in the subdivision tree,
the analysis in \cite{yakoubsohn:bisection-analysis:05} implicitly
assumes that the initial square $B_0$  contains a
\emph{positive} number of zeros $\zeta_i$ with multiplicities $m_i \in
\mathbb{N}$ constituting $p \in \mathbb{N}$ clusters. If we take $p=0$
(or $m_i=0 \, \forall i$) to cover a box devoid of zeros, the
estimates for $Q_{\epsilon}$ (the number of executed exclusion tests)
in ( \cite{yakoubsohn:bisection-analysis:05}, Thms. 1.1 , 1.2) are
obviously wrong.

\end{itemize}


Thus, we hope it is fair to conclude that Yakoubsohn's analysis
essentially (i) deals with root refinement of a positive number of
zeros in a given box, (ii) is placed in the Real RAM setting for
general analytic functions, and (iii) gives a bit-precision estimate
estimate only for square-free polynomials, and expresses it in terms
of the unknown distance to the set of ill-posed problems.


\subsubsection{Comparison with methods based on winding number computations}
Let us consider an analytic function $f$ with no zeros on the boundary
of a compact region $R \subset \mathbb{C}$ whose boundary is given by
a contour (rectifiable Jordan curve) $\gamma$.  Then it is well known
that the number $N$ of zeros of $f$ inside $R$ is given by
$$ s_0:= \frac{1}{2 \pi \sqrt{-1}}\int_{\gamma}\frac{f'(z)}{f(z)} dz.$$

Let the distinct zeros in $R$ be $z_1, \ldots, z_n$ with
multiplicities $\nu_1, \cdots, \nu_n$, and total number $\sum_{i=1}^k
\nu_i = N \in \mathbb{N}_0.$ (Please note the changed indexing of
zeros and their total number in this subsection.) It is well-known
that
$$s_k:= \sum_{i=1}^n z_i^k= \frac{1}{2 \pi \sqrt{-1}}\int_{\gamma}z^k\frac{f'(z)}{f(z)} dz, $$ and that the  Girard-Newton identities may be solved to yield a polynomial with exactly the roots $z_i$ and their multiplicities $\nu_i$.

Delves and Lyness \cite{delves-lyness:roots-analytic-functions:67}
already proposed adaptive subdivision and modification of an initial
region $R=R_0.$ Their subdivision is guided by numerical criteria
capturing smallness of $f$ (indicating zeros near the contour
$\gamma$) whenever $\int_{\gamma} f'(z)/f(z) dz > K$. This together with forcing subdivision whenever
$N=N(R)$ exceeds a threshold $M$ is supposed to lead to
well-conditioned systems of equation. The work of Delves and Lyness does
not come with a computational complexity estimate, and it is not known
whether its numerical strategies actually yield a finite method (an
algorithm).

A different approach to root-approximation with accuracy $\epsilon$
was taken by Ying and Katz \cite{ying-katz:zeros-analyic:88} who use
adaptive quadrature on a polygonal chain around the region $R$ to
determine the integer $s_0$. This might be used as an
inclusion-exclusion test together with subdivision. Their numerical
approach is reliable but may return ``Failure'' (\emph{viz.}
\cite{ying-katz:zeros-analyic:88}, p.157), and its complexity has not
been analyzed.

Another reliable algorithm by Ying and Katz
\cite{ying-katz:simple-reliable-solver:89} uses exclusion predicates
similar to our $C_0$, and/or some  refined version of it, namely: $|f(m(B))| > |f'(m(B))| +
M(B) \cdot r^2/2$, where $M(B) > |\intbox f''(B)|$.  Using either exclusion a
reliable algorithm was proposed which computed an inclusion of the
roots inside the region $R$ up to accuracy $\epsilon$. The run-time
was only analyzed in a Real RAM model, and finite precision issues were
not discussed.

Kravanja, Sakurai and van Barel \cite{Kravanja1999} separate the
determination of the number $n$ of different zeros, and the
approximation of the mutually different zeros $z_1, \ldots, z_n$ from
the computation of their multiplicities. Thereby, ill-conditioning of
root determination is supposedly overcome. Formal ortho\-gonal
polynomials (FOP's) $\phi_0,\phi_1=(z-s_1/s_0), \ldots, \phi_k$ are
computed via contour integrals. The sequence $(\phi_i)_i$ will be terminated with
$k=n$ as the last index of this sequence (so that
$\phi_n=\prod_1^n(z-z_i)$) based on numerical assessment of a sequence
of quadratures. Subsequently, a Vandermonde system connecting the
roots $z_i$ and their multiplicities $\nu_i$ with the symmetric
functions $s_{i-1}$ is solved. Only numerical results have been
reported in \cite{Kravanja1999}. In section 6 \emph{ loc.cit.}  the
authors additionally discuss approximation of all the roots inside a
contour via interpolation at roots of unity connected with multipoint
Pad{\'e} approximants (MPA) (see also \cite{kravanja-vanbarel:bk:00}).
There is no analysis of this variant either.

Derivative-free versions of the approach above have been derived in
\cite{kravanja-vanbarel:derivative-free-zeros-analytic:99}. Also,
approximation properties and numerical stability have been discussed
qualitatively, see
\cite{kravanja+2:error-analysis-derivative-free:2003} . No reliable
verified algorithm, complexity analysis or discussion of finite
precision issues seems to be known.

Dellnitz, Sch{\"u}tze and Zheng
\cite{dellnitz+2:Locating-all-zeros-analytic:2002} propose an adaptive
subdivision \mbox{method} for a function with only simple zeros in a
rectangle $R$. Selection of searched rectangles is driven by winding
number computations relying on adaptive Romberg quadrature. The simple
roots are subsequently approximated via Newton's iteration from the rectangle
center using thresholds as stopping criteria. If verification fails,   rectangles are
halved along the vertical and horizontal bisector alternatingly. No
proof of finiteness or other analysis seems to have been reported, but a combination with convergence criteria for the Newton iteration is
clearly feasible.

Johnson and Tucker
\cite{johnson-tucker:analytic-zeros:09}  suggest to improve enclosures
of a positive number $k$ of zeros. The initial enclosure is found by a winding number
computation, and an improved region is sought  via a Newton iteration with $f^{(k-1)}$. A fixed point of
the iteration may be validated via an interval version of Newton's
method known as the Krawczyck operator. This would eventually allow
verified determination of a zero inside a (``small'') interval box. If
this fails, uniform subdivision is employed. Again, no analysis is provided for the algorithm.

The connection of interpolation to zero-finding was considered again
recently in \cite{austin+2:numerical-algorithms-roots-unity:2014} ,
see also \cite{boyd:bk:chebyshev-proxy-other-rootfinders:2014}, which
latter contains some comparison of methods.
(A reader interested in aspects of fast and stable solving of the
related finite matrix pencils and linear problems may like to consult
\cite{gemignani:zerofinding-structured-matrix:2016}.)
Besides the qualitative fact that zeros of the approximant polynomials
$p_{\mu}$ (of degree $\mu$) converge to the zeros inside the unit
circle when $ \mu \rightarrow \infty$, no quantitative estimates seem
to have been discussed yet.

\section{Preliminary}
\label{sec:prelim}
We borrow some definitions from \cite{yap-sagraloff-sharma:cluster:13,becker+4:cluster:16}. 
Let $f :\CC \mt \CC$ be a holomorphic function and, as above, let $f_k(z)$ be its $k$th normalized derivative,
i.e.,
        $$f_k(z) \as \frac{f^{(k)}(z)}{k!}.$$
The algorithm will work with dyadic numbers $\FF = \set{m2^n: m, n \in \ZZ}$.
A complex number $z$ is represented by a \dt{$\tau$-regular oracle function} $\wt{z}:~\NN~\mt~\FF^2$,
for some fixed $\tau \ge 0$ \cite{becker+4:cluster:16}: Given $L \in \NN$, the oracle returns
$\wt{z}$ such that $|\wt{z} - z| \le 2^{-L}$ and $\wt{z}$ has $O(\tau + L)$ bits. 
 We will often use the notation ``$z$'' to stand
for both the number and its oracle representation. An interval $I \ibp \RR$ is represented
by a pair of oracles for its endpoints.
Let $\intbox \FF$ represent the set of closed intervals with dyadic endpoints.
The algorithm will work with boxes $B$ with dyadic endpoints, i.e.,
$B \in \intbox \FF \times \intbox \FF$. 
Given a function $f: \CC \mt \CC$, a \dt{box-function} for $f$, represented as $\intbox f$,
is a function satisfying the following properties:
\begin{enumerate}
\item $\intbox f: \paren{\intbox \FF}^2 \mt \paren{\intbox \FF}^2$.
\item For all boxes $B$, we have that $f(B) \ibp \intbox f(B)$.
\item The box-function satisfies a first-order error bound, i.e., for a box $B$
        \beql{convergence}
        \sup_{z, w \in \intbox f(B)}\abs{z-w} \le \sup_{z\in B}\abs{f'(z)} \cdot (2w(B)).
        \eeql
\end{enumerate}
The last property ensures  that the box-function is monotonically converging, i.e.,
as $r_B$ tends to zero, the box function converges to the value $f(m_B)$. 
Such error bounds are implemented as part of some interval arithmetic libraries (e.g., \cite{johansson:arb}).
The algorithm asumes that we can compute box-functions for $f$ and all its higher order normalized
derivatives. The error bound for the box-function of the $k$th normalized derivative is
        \beql{convergencek}
        \sup_{z, w\in \intbox f_k(B)}\abs{z-w} \le \sup_{z \in B} \abs{f_{k+1}(z)} \cdot (2(k+1)w(B)).
        \eeql
Box-functions can be computed using interval arithmetic and box-function representations
for some standard class of functions 
are part of multiprecision software libraries \cite{bailey:mpfun,johansson:arb,mpfc,irram,yu+4:core2}.
Given an algorithm computing a box-function for $f$, one can also use techniques from automatic differentiation
to compute a box-function for its derivatives \cite{griewank-walther:bk}. 
Given a disc $D(m,r)$, where $m$ and $r$ are dyadics, and some $k$, by $\intbox f_k(D(m,r))$,
we denote the output of the box function $\intbox f_k(B)$, where $B$ is the smallest axis
aligned box centered at $m$ and containing $D(m,r)$. 


Using the monotonically converging property of box functions, we can also 
construct  oracle functions for the numbers $f_k(w)$, for a dyadic point $w$: Given $w \in \FF^2$,
compute $\intbox f_k(w + 2^{-p}D(0,1))$, for $p = 1, 2, \ldots$, until we obtain the desired level
of accuracy in the output. It is important to note that the complexity analysis later
will not account for the complexity of computing the oracle functions, but only 
bound the precision {\em demanded} from the oracle functions; the $\tau$-regularity ensures
that the number of bits in the output are an additional additive factor of $\tau$ more than the demand.

\subsection{The Soft Zero Test}
\label{sec:soft-zero-test}

As mentioned earlier, a crucial component of the algorithm is the soft zero test. Here we describe
the test for comparing two closed intervals $I$, $J$ in $\RR$. Given a precision $p \in \NN$,
let $(I)_p$ be an interval containing $I$ obtained by computing a $p$-bit approximation to its two endpoints
and then rounding appropriately,
i.e., if $I= [a,b]$ and $\wt{a}$ and $\wt{b}$ are a $p$-bit absolute approximation to $a$ and $b$
respectively, then $(I)_p = [\wt{a}-2^{-p}, \wt{b}+2^{-p}]$. If an interval does not contain the origin, then
its sign is well defined. Given the guarantee that one of the intervals $I$, $J$
does not contain zero, we want to deduce the sign of $I-J$ or deduce that the two intervals
are ``relatively close.'' A procedure resolving this problem is called the soft zero test for intervals.
This is a generalization of similar tests for real numbers given in 
\cite{yap-sagraloff-sharma:cluster:13,becker+4:cluster:16} and is a slight
modification of the comparison operator implemented in \cite{irram}.

\progb{
\texttt{SoftCompare}\\
{\sc Input:} Intervals $I$ and $J$.\\
{\sc Output:} Sign of $I-J$ or that they are ``relatively equal''.\\
1.\> $p = 1$.\\
2.\> do\\
\>\> Compute $(I)_p$ and $(J)_p$.\\
\>\> If $(I)_p \si (J)_p= \es$ then \\
\>\>\> Output the sign of $(I)_p - (J)_p$.\\
\>\> Else if $(I)_p \si (J)_p \neq \es$ and either $0 \nin (I)_p$ or $0 \nin (J)_p$ then\\
\>\>\> Output \texttt{Relatively Close}.\\
\>\> Else $p \ass 2 \times p$.
}
The loop terminates because by assumption at least one of the intervals does not contain zero;
otherwise, in the limit $\set{0} \in I \si J$. Let 
$d(I)$ be the \dt{distance between}
$I$ and the origin: $d(I) \as 0 $ if $0 \in I$; otherwise, 
        $$d(I) \as \min\set{|x|; x \in I}.$$
Similar to \cite{yap-sagraloff-sharma:cluster:13}, we obtain the following
\bleml{softcompare}
In evaluating \texttt{SoftCompare}$(I,J)$, the number of bits bits
requested by the oracles for $I$ and $J$ is bounded by
        $$2-\paren{\log_2\omin(\max\set{d(I),d(J)})}.$$
\eleml
\bpf
Without loss of generality, let us assume $0 \nin I$ and $d(I) \ge d(J)$.
If $I=[a,b]$, $a > 0$, and $(I)_p = [\wt{a}- 2^{-p}, \wt{b}+2^{-p}]$, then $(I)_p$ does not 
contain the origin if $p \ge 1 + \log_2 (1/a)$. Therefore, the iteration will certainly stop 
before the bound mentioned in the statement is attained.
\epf

\section{The Soft Zero Algorithm for Root Clustering}
\label{sec:soft-zero-algo}
For a disc $D(m,r)$, where $m, r$ are dyadic numbers, 
define the predicate 
\beql{wck}
\intbox \wtC_k(m,r) \equiv \texttt{SoftCompare}(\abs{f_k(m)}r^k, \sum_{i=0}^{k-1}\abs{f_i(m)}r^i + \abs{\intbox f_{k+1}(D(m,r))}r^{k+1})>0.
\eeql
The crucial component of the algorithm is the following procedure:
\progb{
\texttt{firstC(B, $N_0$)}\\
{\sc Input:} A box $B \ib \CC$ and a number $N_0$ greater than the number\\
\>\> of roots of $f$ in $B$ counted with multiplicity.\\
{\sc Output:} Smallest $k \le N_0$ such that $D(c_1k\cdot B)$ is isolating; \\
\>\> otherwise return $-1$\\
\> for $k = 0 \dd N_0$ do\\
\>\> If $\intbox\wtC_k(\max\set{1,\paren{c_1k}}\cdot B)$ and $\intbox\wtC_k(\max\set{1,\paren{3c_1k}} \cdot B)$ hold then\\
\>\> \Comment{The max is required to ensure that the box is well-defined for $k=0$.}\\
\>\> \> Return $k$.\\
\> Return -1.
}
The constant $c_1 \as 32e$ appears as a consequence of the converses given in \refSec{conv-incl-excl}
(in particular, from \refLem{geomconv}). 
We will call a box $B$  an \dt{inclusion box}, if the procedure
returns a positive value; it is called an \dt{exclusion box} if it returns zero.

Let the pair $(B, k)$ denote the smallest disc centered at $m(B)$ and containing $c_1kB$.
Two pairs $(B,k)$ and $(B',k')$ are said to be \dt{in conflict} if 
the corresponding discs intersect.
The details of the overall algorithm for the local root clustering problem are as follows:

\progb{
\texttt{Soft Root Clustering Algorithm}\\
{\sc Input:} $f: \CC \mt \CC$, initial box $B_0$, a number $N_0$ upper bounding \\
\>\> the number of roots of $f$ in $B_0$ counted with multiplicity.\\

{\sc Output:} An isolating system $\calD$ for $f$ in $B_0$.\\

\> $Q_0 \ass \set{B_0}$, $Q_1 \ass \es$, $\calD \ass \es$.\\
1.\> while $Q_0$ is not empty do\\
\>\> Remove a box $B$ from $Q_0$.\\
\>\> $k \ass$ \texttt{firstC(B,$N_0$)}.\\
\>\> If $1 \le k \le N_0$, push the pair $(B,k)$ onto $Q_1$.\\
\>\> else if $k= -1$ subdivide $B$ and push its four children onto $Q_0$.\\
\>\> else discard $B$. \Comment{$k=0,$ so there are no roots in $B$.}\\
2.\> while $Q_1$ is not empty do\\
\>\> Remove $(B, k)$ from $Q_1$.\\
\>\> If $(B,k)$ is not in conflict with any other pair in $\calD$ then\\
\>\>\> Push $(B,k)$ onto $\calD$.\\
\> Return $\calD$.
}

\section{Foundational Results}
\label{sec:basic}
In this section, we derive criteria such that if a box $B$ satisfies them then the procedure
\texttt{firstC} will output a non-negative value. 
Let $\calc$ be a cluster of roots of $f$. Recall from equations \refeq{mcalc} and \refeq{rcalc},
respectively, that
$m_\calc$ is its centroid and $r_\calc$ is the radius of the smallest disc centered at $m_\calc$ containing all
the points in $\calc$. 

We first derive tight bounds on the derivatives $f^{(j)}$ when $z$ is sufficiently close to $m_\calc$,
and $r_\calc$ is not very large. 
Let $g$ denote the degree $k \as |\calc|$ polynomial given by
        $$g(z) \as \prod_{\alpha \in \calc} (z-\alpha).$$
Then there is an analytic function $h$ such that
        \beql{fgh}
        f(z) = g(z)h(z)
        \eeql
and $h$ has no roots in $\calc$. 
By Leibniz's rule for differentiation, it follows that for $j\ge 0$
        $$f^{(j)}(z) = \sum_{i=0}^j {j \choose i} g^{(i)} h^{(j-i)},$$
and hence
        \beql{fjz}
        f_j(z) = \sum_{i=0}^j g_{i}(z) h_{j-i}(z).
        \eeql
Since $\deg(g) = k$, note that the terms $g^{(i)}$ for $i > k$ vanish from the right-hand side.
From Walsh's representation theorem \cite[Thm.~3.4.1c]{rahman-schmeisser:bk}, the equation above can be
further simplified to
        \beql{wfjz}
        f_j(z) = \sum_{i=0}^j {k \choose i}(z-\alpha)^{k-i} h_{j-i}(z),
        \eeql
where $\alpha$ depends on $z$ and is contained in $D(m_\calc, r_\calc)$;  here we use the standard
convention that ${k \choose i}=0$, for $i > k$.

Recall from \refeq{gammac} the  definition of the
$\gamma$-function associated with the cluster~$\calc$:
        \beql{gammah}
        \gamma_{\calc}(z)= \gamma(h,z) \as \sup_{k \ge 1} \abs{\frac{h_k(z)}{h(z)}}^{1/k}.
        \eeql
The following claim is based on standard arguments using $\gamma$-functions
\cite{bcss:bk,giusti+2:zeros-analytic:05}:
\bpropl{gammabound}
For an analytic function $h$, if  $z\in \CC$ is such that $h(z) \neq 0$
then for all $w \in \CC$ such that $4|z-w| \gamma(h,z) \le 1$  we have
$\gamma(h,w) \le 4 \gamma(h,z)$.
\epropl
\bpf
Define $u \as |z-w| \gamma(h,z)$; by assumption $u < 1$.
We first claim that 
        \beql{hwhz}
        1-\frac{u}{1-u} \le \abs{\frac{h(w)}{h(z)}} \le \frac{1}{1-u}.
        \eeql
To obtain this, take the Taylor expansion of $h$ around $z$, divide by $h(z)$, and apply triangular
inequality to get
\begin{align*}
\abs{\frac{h(w)}{h(z)}} \le 1 + \sum_{i \ge 1} \gamma(h,z)^i  |z-w|^i = 1 + \sum_{i \ge 1} u^i.
\end{align*}
Since $u < 1$, we can apply the formula for summing geometric
series to obtain
        $$\abs{\frac{h(w)}{h(z)}} \le 1 + \frac{u}{1-u} = \frac{1}{1-u}.$$
The lower bound follows similarly.

Now we have
\begin{align*}
  \abs{\frac{h_k(w)}{h(w)}} 
        &=  \abs{\frac{h(z)}{h(w)}}  \sum_{i \ge 0}\abs{\frac{h_{k+i}(z)}{h(z)}}{k+i \choose i}|z-w|^i\\
        &\le \paren{\frac{1-u}{1-2u}}  \gamma(h,z)^k \sum_{i \ge 0}{k+i \choose i} (\gamma(h,z)|z-w|)^i\\
        &= \paren{\frac{1-u}{1-2u}} \paren{\frac{\gamma(h,z)^k}{(1-u)^{k+1}}}\\
        &= \frac{1}{1-2u}\paren{\frac{\gamma(h,z)}{(1-u)}}^k\\
        &\le 2^{k+1} \gamma(h,z)^k,
\end{align*}
where in the final step we used the fact that $4u \le 1$. Taking the $k$th root on both sides
of the inequality above gives the desired relation.
\epf

We will need the following result later:
\bleml{bound1}
Let $\calc$ be a cluster of $k \ge 1$ roots. If $z \in \CC$ is such that
        $$4k\paren{|z-m_\calc|+r_\calc} \gamma_{\calc}(z) \le 1,$$ 
then
        $$\half< \abs{\frac{f_k(z)}{h(z)}}< \frac{3}{2}.$$
\eleml
\bpf
As in \refeq{wfjz} above, we know that for every $z$ there exists  $\alpha \in D(m_\calc, r_\calc)$ such that
        $$f_k(z) = \sum_{i=0}^k{k \choose i} (z-\alpha)^{k-i} h_{k-i}(z).$$
Pulling out the last term corresponding to $i=k$ we obtain
        $$f_k(z)= h(z)\paren{1 + \sum_{i=0}^{k-1}{k \choose i} (z-\alpha)^{k-i}\frac{h_{k-i}(z)}{h(z)}}.$$
We will derive an upper bound on the absolute value
of the summation term above and show that under the conditions of the lemma it is at most one half.
Let $\gamma_z$ be a shorthand for $\gamma_\calc(z)$.

From the triangular inequality and  the definition of $\gamma(h,z)$, we obtain
        \beql{bd1}
        \abs{\sum_{i=0}^{k-1}{k \choose i} (z-\alpha)^{k-i}\frac{h_{k-i}(z)}{h(z)}}\le \sum_{i=0}^{k-1}{k \choose i}|z-\alpha|^{k-i}\gamma_z^{k-i}.
        \eeql
Since $\alpha \in D(m_\calc, r_\calc)$, it follows that
        $$|z-\alpha| \le |z-m_\calc| + r_\calc \ras \Delta.$$
Substituting this upper bound in the right-hand side of \refeq{bd1}, we get
\begin{align*}
  \abs{\sum_{i=0}^{k-1}{k \choose i} (z-\alpha)^{k-i}\frac{h_{k-i}(z)}{h(z)}}
                \le \sum_{i=0}^{k-1}{k \choose i} (\Delta\gamma_z)^{k-i}
                = {(1+\Delta\gamma_z)^{k}-1}.  
\end{align*}
The condition in the statement of the lemma states that
        $4k\Delta \gamma_z \le 1$,
which implies that $(1+\Delta \gamma_z)^k < e^{1/4} < 3/2$, and hence the desired result follows. \epf

As a corollary, it follows that if $0<2kr_\calc \gamma_\calc(m_\calc) \le 1$ then
 $f_k(m_\calc) \neq 0$; in fact, there is a small neighborhood around $m_\calc$ where
$f_k$ does not vanish. The result above states that when $z$ is sufficiently close to the
cluster $\calc$ and $r_\calc$ is sufficiently small then $|f_k(z)|$ and $|h(z)|$ are almost 
similar in value. The next result states that if $z$ is sufficiently close to $m_\calc$ then
$|f_k(z)|$ and $|f_k(m_\calc)|$ are also similar in value.
\bleml{fkzbd}
If $\calc$ is a cluster of size $k \ge 1$, 
then for all $z$ such that
        $$4k|z-m_\calc| \gamma_\calc(m_\calc) \le 1,$$
we have  
        $$\half < \abs{\frac{f_k(z)}{f_k(m_\calc)}} < \frac{3}{2}.$$
\elem
\bpf
We first prove the lower bound.
From Taylor expansion of $f_k(z)$ around $m_\calc$ we obtain that
        $$\abs{f_k(z)} \ge |f_k(m_\calc)| \paren{2 - \sum_{i \ge 0} {k+i \choose i} u^i}$$
where $u \as \gamma_\calc(m_\calc) |z-m_\calc|$.
The summation on the right-hand side above is equal  to $(1-u)^{-k-1}$. Therefore,
        $$\abs{f_k(z)} \ge |f_k(m_\calc)| \paren{2 - \frac{1}{(1-u)^{k+1}}}.$$
Since $4ku<1$, by assumption, 
a simple numerical argument shows that
$(1-u)^{-(k+1)} < 3/2$, for all $k \ge 1$, which gives the desired lower bound.

The upper  bound follows similarly from the inequality that
        $$\abs{f_k(z)} \le |f_k(m_\calc)| \paren{\sum_{i \ge 0} {k+i \choose i} u^i} 
                                 = |f_k(m_\calc)| \frac{1}{(1-u)^{k+1}}.$$

\epf

\subsection{Growth estimates on a disc} %
\label{sec:estim-maxima-disc}

Let
        \beql{Mf}
        M(f,k,m,r) \as \sup_{w \in D(m,r)} \abs{f_k(w)} = \sup_{w \in \partial D(m,r)} \abs {f_k(w)}
        \eeql
be the upper bound on the absolute value of $f_{k}(D(m,r))$, for $k \ge 0$; we will
simply write $M(f,m,r)$ when $k=0$, and use the notation 
$M_{k,m,r}$ when $f$ is understood from the context. 
We will frequently need estimates and relations on $M(f,k,m,r)$.

From Cauchy's differentiation formula \cite{tichmarsh:function:bk} we know that for $j\ge k$
        \beql{cauchy}
        {j \choose k} f_{j}(m) = \frac{1}{2\pi i} \oint_{D(m, r)} \frac{f_{k}(z)}{(z-m)^{j-k+1}} dz.
        \eeql
We derive a few useful consequences of this result. 
\begin{enumerate}
\item Taking the absolute values and applying triangular inequality, it follows that for
  a disc $D(m,r)$, we have
        \beql{mflb}
        M(f, k+1, m,r) \ge r^{j-k} |f_j(m)| {j \choose k+1}.
        \eeql
\item Let $w \in D(m,r)$ be a point attaining $M(f,k+1,m,r)$. Then applying \refeq{cauchy} 
  with $j=k+1$ to the disc $D(w,r) \ibp D(m,2r)$ we obtain
        \beql{mfk1}
        (k+1)M(f,k+1,m,r) \le  \frac{M(f,k,m,2r)}{ r}.
        \eeql
\end{enumerate}

The next lemma is an estimate on the largest absolute value in $\intbox f_k(B)$ in terms of
$M(f,k,2B)$, and will be useful later in deriving converses.
\bleml{boxbound}
Given a box $B$, and $k \ge 0$, the convergence condition \refeq{convergencek} implies that
        \beql{overest}
        \sup_{z \in \intbox f_k(B)} |z| < 3 M(f,k,m_B,2r_B).
        \eeql
\eleml
\bpf
From \refeq{convergencek} and the bound in \refeq{mfk1} we obtain that
\begin{align*}
  \sup_{z, w\in \intbox f_k(B)}|z-w| 
                \le 2r_B (k+1) M_{k+1, m_B, r_B}
                \le 2M_{k,m_B,2r_B}.
\end{align*}

Let $z'\in \intbox f_k(B)$ attain the largest absolute value in the box-function, and  let $w' \in f_k(B)$  be similarly
defined. Since $\intbox f_k(B)$ contains $f_K(B)$, it also contains  $w'$, and hence combined with the bound above
        $$|z'| - |w'| \le |z'-w'| \le 2M_{k, m_B, 2r_B}.$$
Since $f$ is holomorphic, from the maximum modulus principle we know that
$M_{k, m_B, r_B} \le M_{k, m_B, 2r_B}$, whence  the desired claim.

\epf

We will also need the following upper bound on $M(f,k,m,r)$ near a cluster $\calc$.
\bleml{mfbnd}
Let $\calc$ be a cluster of $k \ge 1$ roots. If  $m$ and $r$ are such that
        $$4k(|m-m_\calc|+r_\calc+r) \gamma_\calc(m) \le 1,$$
then 
        $$M(f, k+1, m,r) \le 16e\gamma_\calc(m) |f_k(m)|.$$
\eleml
\bpf
Let $w$ be a point on the boundary of $D(m,r)$ where $M_{k+1, m,r}$ is attained.
From \refeq{wfjz} we know that there is an $\alpha \in D(m_\calc, r_\calc)$ such that 
        $$f_{k+1}(w) = h'(w)  + \sum_{j=1}^{k} {k \choose j} (w-\alpha)^jh_{j+1}.$$
Since $4k(|m-m_\calc|+r_\calc) \gamma_{\calc}(m) \le 1$, 
we use the lower bound  on $|f_k(m)|$ in terms of $|h(m)|$  from  \refLem{bound1} to obtain
        $$\abs{\frac{f_{k+1}(w)}{f_k(m)}} \le \frac{2}{|h(w)|}\paren{\abs{{h'(w)}}  + \sum_{j=1}^{k} {k \choose j}|w-\alpha|^j\abs{{h_{j+1}(w)}}}.$$
Dividing and multiplying the right-hand side by $|h(m)|$ we obtain
        $$\abs{\frac{f_{k+1}(w)}{f_k(m)}} \le 2\frac{|h(m)|}{|h(w)|}\paren{\abs{\frac{h'(w)}{h(m)}}  + \sum_{j=1}^{k} {k \choose j}|w-\alpha|^j\abs{\frac{h_{j+1}(w)}{h(m)}}}.$$
Since $4r\gamma_{\calc}(m) \le 1$, we can use the relation between
$|h(w)|$ and $|h(m)|$ from \refeq{hwhz}  along with the definition of $\gamma_\calc(w)$ to get
        $$\abs{\frac{f_{k+1}(w)}{f_k(m)}} \le 4\paren{\gamma_w  + \sum_{j=1}^{k} {k\choose j}|w-\alpha|^j \gamma_\calc(w)^{j+1}}.$$
Since $w$ is on the boundary of $D(m,r)$ and $\alpha \in D(m_\calc, r_\calc)$, it follows that 
        $$|w-\alpha| \le |w-m| + |m-m_\calc|+ r_\calc = r + |m-m_\calc| + r_\calc \ras \Delta.$$
Therefore,
        \beql{fk1m}
        \abs{\frac{f_{k+1}(w)}{f_k(m)}} \le 4\gamma_\calc(w) (1+ \Delta \gamma_\calc(w))^k.
        \eeql
Since $4r\gamma_\calc(m) \le 1$, we know from \refPro{gammabound} that
$\gamma_\calc(w) \le 4\gamma_\calc(m)$, and hence the 
statement of the lemma implies that $k\Delta\gamma_{\calc}(w) \le 1$.  Therefore, the term
in the parenthesis above is at most $e$, giving us the desired upper bound.
\epf

\subsection{Relation between $\gamma(f,z)$ and $\gamma_\calc(z)$}
\label{sec:relat-betw-gamm}
Let $\calc$ be a cluster of $k \ge 1$ roots and $g(z)$, $h(z)$ be as in \refeq{fgh}. For $z \in \CC$, define
        \beql{sgz}
        S_g(z) \as \sum_{\alpha \in \calc} \frac{1}{|z-\alpha|}.
        \eeql
From \refeq{fjz} we know that
        $$f_j(z) = \sum_{i=0}^j g_i(z) h_{j-i}(z)$$
where the terms $g_i$ vanish for $i > |\calc|$. Then we have that
        $$\abs{\frac{f_j(z)}{f(z)}} = \abs{\sum_{i=0}^j \frac{g_i(z)}{g(z)}\cdot \frac{h_{j-i}(z)}{h(z)}}
                                                     \le {\sum_{i=0}^j \abs{\frac{g_i(z)}{g(z)}}\cdot \abs{\frac{h_{j-i}(z)}{h(z)}}}.$$
From the inequality $|g_i(z)/g(z)| \le S_g(z)^i$, which is trivially true if $i > |\calc|$,
and the definition of the  function $\gamma_\calc$ it follows
that 
        $$\abs{\frac{f_j(z)}{f(z)}} \le \sum_{i=0}^j S_g(z)^i \gamma_\calc(z)^{j-i} 
                                                < (S_g(z) + \gamma_\calc(z))^j.$$        
Taking the $j$th root on both sides, we obtain that for all $z\in \calc$
        \beql{gammaub}
        \gamma(f,z) < S_g(z) + \gamma_\calc(z).
        \eeql
Since the argument never relied on the actual definition of a cluster, it works even when $\calc$ is any
multiset of roots of $f$ with appropriate definitions of $g$ and $h$.

\subsection{Geometric Interpretation of $\gamma(f,z)$}
\label{sec:geom-interpr}
The function $\gamma(f,z)$ upper bounds the growth of the 
absolute value of the higher order Taylor coefficients  $f_k(z)$ at $z$ with respect to the 
absolute value of the first term $f(z)$, which is assumed to be non-zero. 
It is not a priori clear that this can be uniformly
bounded; e.g., the first two derivatives of the function $e^{e^{z}}$ are $e^z e^{e^z}$
and  $e^z e^{e^z} + e^{2z} e^{e^z}$, which grow much faster than the function itself; 
this issue does not arise in the case of polynomials. 
However, from de Branges's theorem we know that if the function is ``nice''
at a point then the $k$th Taylor coefficient of the function is bounded by at most $k$.
To formalize this notion of
niceness we need the following definition:
A function $h: D \mt \CC$, on a domain $D \ibp \CC$, is called \dt{univalent} (or schlicht)
on $D$ if it is injective (i.e, one-to-one).
The concept of niceness is the following: 
A function $h: D \mt \CC$ that is {\em holomorphic and univalent} on $D$ is called a \dt{conformal mapping} of
$D$, or conformal on $D$.
Analogous to the real analytic setting, it is well known \cite[p.~198--200]{tichmarsh:function:bk} 
that the non-vanishing of the derivative of a holomorphic function at a point is a necessary and 
sufficient condition  to ensure that the function is conformal on some neighborhood of the point.
The de Branges's theorem \cite{conway:complex2:95} states that the Taylor coefficients   of a conformal map on a disc
possess (and eventually even  attain) a sharp bound.

\bthmT{de Branges's Theorem}{dbt}
Let $h$ be a  conformal map on the open unit disc with Taylor series of the form
        $$h(z)=z+\sum_{n\geq 2} h_n(0) z^n,$$
i.e., $h(0)=0$ and $h'(0) =1$. Then for all $n \ge 2$,
        $\abs{h_n(0)} \leq n.$
\ethmT
We first formulate a non-normalized or extended version of the theorem above:
\bthmT{Extended de Branges's Theorem}{edbt}
Let $g$ be a  conformal map on the disc $D(\alpha, R)$, where $\alpha$ is a root of $g$
such that $g'(\alpha) \neq 0$. Then for all $n \ge 2$,
        $$\abs{\frac{g_n(\alpha)}{g'(\alpha)}} \leq \frac{n}{R^{n-1}}.$$
\ethmT
\bpf
Consider the function $h(z) \as g(\alpha + Rz)/(Rg'(\alpha))$. Then every point~$z$ in
the unit disc $\mathbb{D}$ is mapped to some point in $D(\alpha, r)$. In particular,
$h(0) = g(\alpha)=0$ and $h'(0) = 1$. Further note that $h_n(0) = R^{n-1}g_n(\alpha)/g'(\alpha)$.
Clearly, $h$ is conformal on the unit disc and hence from \refThm{dbt} we get that
        $$n \ge \abs{h_n(0)} = R^{n-1}\abs{\frac{g_n(\alpha)}{g'(\alpha)}}$$
which gives us the desired upper bound.

\epf

How do we apply this theorem in our context? Consider an analytic function $f$ 
such that  $f(\alpha) \neq 0$, and  define the function $g(z) \as (z-\alpha)f(z)$.
Since $g'(\alpha) = f(\alpha) \neq 0$, we know that there is a disc around $\alpha$
on which $g$ is conformal. 
Define \dt{$R_f(\alpha)$ as the largest radius around $\alpha$ on which $g$ is conformal}. 
Then we can inductively verify that
        $$g^{(n)}(z) = nf^{(n-1)}(z) + (z-\alpha) f^{(n)}(z).$$
Therefore, $g_{n+1}(\alpha) = f_{n}(\alpha)$ for $n \ge 0$, which implies that
        \beql{gammaradius}
        \gamma(f,\alpha) = \sup_{n \ge 1}\abs{\frac{f_n(\alpha)}{f(\alpha)}}^{1/n} 
                              = \sup_{n \ge 1}\abs{\frac{g_{n+1}(\alpha)}{g'(\alpha)}}^{1/n} 
                              \le \frac{\sup_n (n+1)^{1/n}}{R_f(\alpha)} < \frac{e}{R_f(\alpha)},
        \eeql
where the penultimate inequality follows from \refThm{edbt}.

A quantitative way to derive a lower bound on the radius of univalence is facilitated by  the following result
\cite{Harris1977}:
\bthmT{Bloch's theorem}{lt}
Let $h: D(\alpha, r) \mt \CC$ be a holomorphic function such that $h(\alpha)=0$, $h'(\alpha) \neq 0$
and for all $z \in D(\alpha, r)$, $|h'(z)| \le M$, for some $M$.  Define 
        $$R \as \frac{r|h'(\alpha)|}{M} \text{ and } R' \as \frac{r|h'(\alpha)|}{2M}.$$
Then $h$ maps $D(\alpha, R)$ biholomorphically onto $D(0, R')$.
\ethmT

When is $R$ maximized? There is a trade-off between the increasing value of $r$ and $M$, since $M$ also
increases with $r$. 
So, it makes sense to define 
        $$r_h(\alpha) \as \sup_r \frac{r}{\sup_{z\in D(\alpha,r)}|h'(z)|}.$$
Moreover, from Cauchy's differentiation formula \refeq{cauchy} it follows that for all $z \in D(\alpha, r)$
        $$|h'(z)| \le \frac{\sup_{z \in D(\alpha,2r)}|h(z)|}{r}.$$
Therefore, $r_h(\alpha) \ge r^2/M(h,\alpha, 2r)$, for all $r \ge 0$, where 
$M(h, \alpha, r) \as \sup_{z \in D(\alpha,r)}|h(z)|$. Substituting $h \as (z-\alpha)f$ we 
obtain that for all $r \ge 0$,
        \beql{roulbd}
        R_f(\alpha) \ge |f(\alpha)| r_h(\alpha) \ge |f(\alpha)|\frac{r}{2M(f,\alpha,2r)}.
        \eeql
Combining this with \refeq{gammaradius} we get that if
$f(\alpha) \neq 0$ then
        \beql{gammafbd}
        \gamma(f,\alpha) = O\paren{\frac{M(f,\alpha, 2r)}{r|f(\alpha)|}}
        \eeql
for all $r \ge 0$.

We use the results above to  derive a similar bound on $\gamma_\calc(m_\calc)$ (defined in \refeq{gammac}),
where $\calc$ is a strongly separated cluster. In particular, this implies that 
$c_0 |\calc|^2 r_\calc \gamma_\calc(m_\calc) \le 1$, therefore, from \refLem{bound1} we get that 
$|f_k(m_\calc)| = \Theta(h(m_\calc))$ and both are non-zero. Combining this 
result with the bound in~\refeq{gammafbd} with $f$ as $h$ and $\alpha$ as $m_\calc$, 
we directly obtain that
        $$\gamma_\calc(m_\calc) = O\paren{\frac{M(h,m_\calc, 2r)}{r|f_k(m_\calc)|}}$$
for all $r \ge 0$. It only remains to upper bound $M(h, m_\calc, 2r)$ in terms of $f$. 
Since $f = gh$, it follows that for all $z$ such that $r\as |z - m_\calc| \ge 2r_\calc$, we have
$|g(z)|\ge (r/2)^k$. Therefore, we obtain the following bound: If $\calc$
is a strongly separated cluster then
        \beql{gammacbound}
        \gamma_\calc(m_\calc)  = O\paren{\frac{2^k M(f,m_\calc,2r)}{r^{k+1}|f_k(m_\calc)|}}
        \eeql
for all $r \ge 2 r_\calc$.

\section{Converse for Inclusion and Exclusion}
\label{sec:conv-incl-excl}

We now use the result above to derive a converse for the predicate $\intbox \wtC_k(m,r)$ 
defined  in \refeq{wck}.

From \cite[Lemma 5.2(b)]{yap-sagraloff-sharma:cluster:13}, we know that  $\intbox \wtC_k(m,r)$ 
holds true if
        $$
        \abs{f_k(m)} r^k >2 \paren{ \sum_{i=0}^{k-1} |f_i(m)|r^i 
                                                        +\intbox f_{k+1}(D(m,sr)) r^{k+1}},
        $$
From  \refLem{boxbound}
we know that the  predicate above holds if the following stronger claim is true:
        \beql{pred}
        \abs{f_k(m)} r^k > 2 \paren{ \sum_{i=0}^{k-1} |f_i(m)|r^i + 
                3 M\paren{f,k+1,m,2r} r^{k+1}}.
        \eeql
This claim follows from the following two claims:  
        \beql{pred1}
        \abs{f_k(m)} r^k > 4 { \sum_{i=0}^{k-1} |f_i(m)|r^i}
        \eeql
and
        \beql{pred2}
        \abs{f_k(m)} > 12 M\paren{f,k+1,m,2r}r.
        \eeql
In the next two sections, we derive conditions for these two claims to hold when $k \ge 1$,
which will give a converse for inclusion;
in \refSec{exclusion-case}, we derive conditions for the special case when $k=0$, which
will give a converse for exclusion;
the results in \refSec{upper-bound-less-k}, \refSec{upper-bound-greater-k}
and \refSec{exclusion-case} are combined to given an overall converse in \refSec{overall-converse}.

\subsection{Upper bound on the sum of derivatives up to $k$}
\label{sec:upper-bound-less-k}
In this section we derive conditions for \refeq{pred1} to hold near a cluster $\calc$
of $k \ge 1$ roots. The conditions, intuitively speaking, should
be that $m$ is close to $m_\calc$, the centroid of the cluster, and $r$ is sufficiently large to include
$D(m_\calc, r_\calc)$. 

Moving the term on the left-hand side of \refeq{pred1} to the right-hand side, we would like to 
show that 
        \beql{cond1}
        1 \ge K\sum_{j=0}^{k-1}\abs{\frac{f_j(m)}{f_k(m)}}r^{j-k},
        \eeql
for $K= 4$.
We next derive a sequence of upper bounds on the right-hand side 
under some suitable assumptions.

Define 
        $$\Delta \as |m-m_\calc|+r_\calc \ge |m-\alpha|,$$
for all $\alpha \in D(m_\calc, r_\calc)$.
Let $\gamma_m$ be a shorthand for $\gamma_\calc(m)$, defined in  \refeq{gammah}.
Suppose $m$ is such that
        \beql{condm}
        4k\Delta \gamma_m \le 1
        \eeql
then we can substitute the lower bound on $|f_k(m)|$ from  \refLem{bound1},
to  obtain the following:
        $$K\sum_{j=0}^{k-1}\abs{\frac{f_j(m)}{f_k(m)}}r^{j-k}\le 2K \sum_{j=0}^{k-1}\abs{\frac{f_j(m)}{h(m)}}r^{j-k}.$$
From the expression for $f_j(z)$ in \refeq{wfjz} and the triangular inequality, 
we obtain that there is an $\alpha \in D(m_\calc, r_\calc)$ 
such that the quantity on the right-hand side above is not larger than
        $$
        \begin{aligned}[b]
        &2K\sum_{j=0}^{k-1}r^{j-k} \sum_{i=0}^j {k \choose i} |m-\alpha|^{k-i} \abs{\frac{h_{j-i}(m)}{h(m)}}\\
         &\quad \le 2K\sum_{j=0}^{k-1}r^{j-k} \sum_{i=0}^j {k \choose i} \Delta^{k-i} \gamma_m^{j-i},
        \end{aligned}
        $$
where the inequality follows from the definitions of $\Delta$, given above, and
the $\gamma$-function \refeq{gammah} for $h$. Since $k > j$, we can write $\Delta^{k-i}$
as $\Delta^{k-j+j-i}$ and pull out $\Delta^{k-j}$ from the inner summation to get
the following form for the right-hand side above
        $$2K\sum_{j=0}^{k-1}\paren{\frac{\Delta}{r}}^{k-j} \sum_{i=0}^j {k \choose i} (\Delta\gamma_m)^{j-i}.$$
Further pulling out $k!$ from the inner summation, and multiplying and dividing by $j!(k-1-j)!$, we get
that the quantity above is equal  to 
        \beql{aux2}
        2Kk\sum_{j=0}^{k-1}{k-1 \choose j}\paren{\frac{\Delta}{r}}^{k-j} 
                \sum_{i=0}^j \frac{j!(k-1-j)!}{i!(k-i)!} (\Delta\gamma_m)^{j-i}.
         \eeql
Since $i \le j$, we have 
        $$\frac{(k-i)!}{(k-1-j)!} = (k-i) (k-i-1) \ldots (k-j)$$
which is greater than $(j-i)!$, since $k > j$. Substituting this bound in \refeq{aux2},
we have so far shown that 
        $$
        \begin{aligned}
          K\sum_{j=0}^{k-1}\abs{\frac{f_j(m)}{f_k(m)}}r^{j-k}
          &\le 2Kk\sum_{j=0}^{k-1}{k-1 \choose j}\paren{\frac{\Delta}{r}}^{k-j}\sum_{i=0}^j {j \choose i} (\Delta \gamma_m)^{j-i}\\
          &= 2Kk\sum_{j=0}^{k-1}{k-1 \choose j}\paren{\frac{\Delta}{r}}^{k-j} (1+\Delta \gamma_m)^j\\
          &= \frac{2kK\Delta}{r}\sum_{j=0}^{k-1}{k-1 \choose j}\paren{\frac{\Delta}{sr}}^{k-1-j} (1+\Delta \gamma_m)^j\\
          &= \frac{2kK\Delta}{r}\paren{\frac{\Delta}{r}+1+\Delta \gamma_m}^{k-1}.
        \end{aligned}
        $$
Assuming $r \ge 2 k\Delta$, along with the condition in \refeq{condm}, we obtain that
the term in the parenthesis above is at most $(1+1/k)^{k-1}$, which is at most
$e$. Therefore, to ensure \refeq{cond1},
it suffices that  $r \ge 2kKe\Delta$. If we further assume that $2r \gamma_m \le 1$, then we
obtain \refeq{condm} as a consequence. 
To summarize, we have the following result, substituting $K=4$:
\bleml{bound2}
Let $\calc$ be a cluster of $k\ge 1$ roots. 
If $m\in \CC$ and $r \in \RR_{\ge 0}$ are such that
        $$8ek\paren{|m-m_\calc|+ r_\calc} \le r \le \frac{1}{2\gamma_\calc(m)}$$
then
        $$\abs{f_k(m)}r^{k}\ge 4\sum_{j=0}^{k-1}\abs{f_j(m)}r^j.$$
\eleml

Note that the conditions in the lemma above are stronger than those in \refLem{bound1}.

\subsection{Upper bound on the tail term}
\label{sec:upper-bound-greater-k}
In this section, we derive conditions for \refeq{pred2} similar to the result in \refLem{bound2}.
Since the condition in some way depends on roots outside a cluster $\calc$ of $k \ge 1$ roots, 
there will be a slight difference with \refLem{bound2}, in that we will require a stronger upper bound on $r$.
Recall the shorthand notation  $M_{k+i,m,r}$ for $M(f, k+i, m,r)$, $i=0,1$,  from \refeq{Mf}.

Assuming 
        $$
        4k(|m-m_\calc|+r_\calc+2r)\gamma_\calc(m) \le 1,
        $$
we can substitute
the upper bound from \refLem{mfbnd} on $M_{k+1,m,2r}$ in the right-hand side of \refeq{pred2}
to get the following stronger condition:
        $$
        1 \ge 2^8e \gamma_\calc(m) r.
        $$
If we additionally assume that $r \ge 4k(|m-m_\calc|+r_\calc)$ then 
        $$4k(|m-m_\calc|+r_\calc+2r) \le (8k+1)r \le 9kr,$$
since $k \ge 1$.  Therefore, we get the following:
\bleml{bound3}
Let $\calc$ be a cluster of $k\ge 1$ roots. If $m \in \CC$, and $r \in \RR_{\ge 0}$ are such that
        $$4k \paren{|m-m_\calc|+r_\calc} \le r \le \frac{1}{2^8ek \gamma_\calc(m)}$$
then \refeq{pred1} holds.
\eleml

\subsection{The Exclusion Case $k=0$}
\label{sec:exclusion-case}
The case $k=0$ needs special attention  for getting a complete converse to  $\texttt{firstC}$;
the arguments in previous sections assumed $k \geq  1$ and hence do not apply. 
In this case, the inequality \refeq{pred} for $k=0$ is
        \beql{excl}
        |f(m)| \ge 6 \cdot M_{1, m,2r} \cdot r.
        \eeql
Let $w$ be a point on the boundary of $D(m,2r)$ where the supremum of the
first derivative is achieved. Recall the definition of $\gamma(f,z)$:
        \beql{gammaz}
        \gamma(f, z) \as \sup_{k \ge 1} \abs{\frac{f_k(z)}{f(z)}}^{1/k}.
        \eeql
We claim  that if $8r\gamma(f,w)  \le 1$ then $4|f(m)| \ge |f(w)|$: From Taylor expansion
around $w$, we observe that
        $$|f(m)| \ge |f(w) |\paren{1 - \sum_{i \ge 1} (2r\gamma(f,w))^i}
                        = |f(w) |\paren{2 - \frac{1} {1-2r\gamma(f,w)}} \ge \frac{|f(w)|}{4}.$$
Hence
        $$\frac{M_{1, m, 2r }}{|f(m)|} \le 4 \abs{\frac{f'(w)}{f(w)}} \le  4\gamma(f,w).$$
Therefore, in order for \refeq{excl} to hold, it suffices that
        $$1 \ge 24r \gamma(f,w).$$
Furthermore, if there exists a $z \in D(m,r)$ such that 
$4r \gamma(f,z) < 1$ then from \refPro{gammabound} we know 
that $\gamma(f,w) \le 4 \gamma(f,z)$, which gives us the following:
\bleml{excl}
If there exists a $z\in D(m,r)$ such that $c_3\gamma(f,z) r \le 1$,
where $c_3 \as 2^7$, then the exclusion test will be successful, i.e., 
\texttt{firstC} will return zero.
\eleml

\subsection{Overall Converse for Strongly Separated Clusters}
\label{sec:overall-converse}

Combining  \refLem{bound2} and \refLem{bound3}, 
we obtain the following result:
\bleml{term}
Let $\calc$ be a cluster of $| \calc|=k\ge 1$ roots. If $m \in \CC$ and $r \in \RR_{\ge 0}$ are such that
        $$8e k \paren{|m-m_\calc|+r_\calc} \le r \le \frac{1}{2^8ek \gamma_\calc(m)},$$
then $\intbox \wt{C}_k(m,r)$ holds.
\eleml

Recall the definitions of $R_\calc$ \refeq{Rcalc} and  a strongly separated cluster  \refeq{ssc} which required
that $r_\calc \le R_\calc/(8|\calc|^3)$.
The gap between $r_\calc$ and $R_\calc$ for a strongly-separated
cluster ensures that if 
the radius $r$ of a disc $D(m,r)$ is in this gap and $m$ is in a neighborhood of $\calc$ then 
we should be able to detect the cluster. More formally:
\bleml{geomconv}
Let $\calc$ be a strongly separated cluster of $k>0$ roots. If $D(m,r)$ is such that
        $$\frac{|m-m_\calc| + r_\calc}{4} \le r \le \frac{R_\calc}{|\calc|^3}$$
then for all $s \in [c_1|\calc|, 3c_1|\calc|]$, where $c_1 \as 32e$,
the predicate $\intbox \wt{C}_k(m,sr)$ holds.
\eleml
\bpf
Validity of the lower bound is straightforward, since for the smaller 
choice of $s$, we have $sr \ge 8e|\calc|(|m-m_\calc| + r_\calc)$, which is the lower bound
in \refLem{term}. Taking the larger value of $s$ and combining with 
the upper bound on $r$ we see that 
        $$3c_1|\calc| r < 3c_1 |\calc|\frac{R_\calc}{|\calc|^3}
                                    = \frac{3c_1}{c_0 |\calc|\gamma_\calc(m_\calc)} 
                                    \leq \frac{1}{2^8 e|\calc| \gamma_\calc(m_\calc)} 
                                    \le \frac{1}{64e|\calc|\gamma_\calc(m)},$$
since $|m-m_\calc|\le R_\calc/|\calc|^3\le 2/(c_0\gamma_\calc(m_\calc))$, 
it follows from \refPro{gammabound} that $\gamma_\calc(m) \le 4\gamma_\calc(m_\calc)$, 
which explains the last inequality. 
Therefore, the claim follows from \refLem{term} by substituting $sr$ for $r$.
\epf


\section{Bounding the Size of the Subdivision Tree}\label{sec:integral-bound}
In this section we derive an upper bound for the size of the subdivision tree (on the input box $B_0$) as created by the 
\texttt{Soft Root Clustering} algorithm,
given in \refSec{soft-zero-algo}.
As a step in that direction, we first refine this subdivision tree as follows.
For every leaf in the subdivision tree where a cluster was detected, we continue the subdivision
process until we either detect a strongly separated cluster or the exclusion test holds.
Let $T_0$ be the tree so obtained. Clearly, the size of $T_0$ is greater than the size of the original tree,
and so it suffices to bound the size of $T_0$.

Recall the definition of the set $S_0$ from \refeq{s0}:
        $$
        S_0= \set{\text{maximal strongly separated clusters $\calc$ partitioning the roots of $f$ in $2B_0$}}.
        $$
From the definition it follows that for a  $\calc \in S_0$ we have $D(m_\calc, r_\calc) \ib 2B_0$,
which implies that $|m_{B_0} - m_\calc|+r_\calc \le 2r_{B_0}$.
If additionally $\calc$ is such that $r_{B_0}~\le~R_\calc/|\calc|^3$ then from 
\refLem{geomconv} we know that \texttt{firstC} will
detect $\calc$ at $B_0$ and we will terminate right away with the size of $T_0$ as one. 
Therefore, from now on we make the following assumption without loss of generality:
        \beql{assum}
          \text{{\bf For all $\calc \in S_0$, $r_{B_0} > \frac{R_{\calc}}{|\calc|^3}$.}}
        \eeql
The next lemma claims that all the clusters detected in $T_0$ are  in $S_0$. 

\bleml{widthinc}
Let $\calc \in S_0$. If $B$ is the first box depth-wise in $T_0$ such that 
$B \si D_\calc/3 \neq \es$ and $B \ibp D_\calc$ then \texttt{firstC} invoked on $B$ detects $\calc$.
\eleml
\bpf
We will use the shorthand $R \as R_\calc/|\calc|^3$.
Let $B'$ be the parent of $B$. Therefore, $B' \si D_\calc/3\neq \es$, but $B' \not \subseteq D_\calc$.
If $p \in B \si D_\calc/3$, then 
this means that $D(p, 4 r_B)$ is not contained in $D_\calc$, i.e.,
        $$|m_\calc - p| + 4 r_B \ge R.$$
But as $p \in D_\calc/3$, we know that $|m_\calc-p| < R/3$. Combining it with the inequality above
we obtain that
        $$r_B \ge \frac{R}{6}.$$
Further recall that for a strongly separated cluster, $r_\calc \le R/8$, which combined with
the inequality above implies that $r_\calc \le 3r_B/4$.
Substituting these bounds we obtain that
        $$|m_B-m_\calc|+r_\calc \le |m_\calc-p| + |p-m_B| +r_\calc \le  \frac{R}{3} + r_B + \frac{3}{4}r_B\le \frac{15}{4} r_B < 4 r_B,$$
which is the desired lower bound in \refLem{geomconv}. The upper bound trivially follows
since $B \ibp D_\calc$.

Note that $B$ cannot be equal to $B_0$ because by \refeq{assum} 
$r_{B_0} > R_\calc/|\calc|^3$ for all $\calc \in S_0$,
therefore, the parent $B'$ of $B$ always exists.
\epf

The lemma above shows that when the radius $r_B$ is somewhat proportional to $R_\calc/|\calc|^3$
and $B$ is in $D_\calc$ then \texttt{firstC} invoked on $B$ detects $\calc$. An area argument shows
that there are  at most a constant number of leaf-boxes covering $D_\calc/3$ that will
detect a $\calc \in S_0$ in such a manner. Note that if $D_\calc/3$ does not intersect $B_0$
then it will not be detected, therefore, we can refine the statement \refeq{assum} on clusters
in $S_0$ to make the following assumption:
        \beql{assum1}
          \text{{\bf For all $\calc \in S_0$, $r_{B_0} > \frac{R_{\calc}}{|\calc|^3}$ and $D_\calc/3 \si B_0\neq \es$.}}        
        \eeql
To account for all the other leaf-boxes $B$, including those that detect a strongly separated cluster,
we observe first that at their parent $B'$ the exclusion test failed. Therefore, from \refLem{excl}, we know
that for all $z \in B'$, $c_3\gamma(f,z) \ge 1$, which is equivalent to $(c_3\gamma(f,z))^2 \ge 1$.
Since the area of $B'$ is four times the area of $B$,  we deduce that
for all such boxes 
$$4\int_B c_3^2\gamma(f,z)^2 \da \ge 1.$$
Combining this with the constant number of boxes covering $D_\calc/3$ that detect a $\calc \in S_0$,
along with the geometric upper bound \refeq{gammaradius} on $\gamma(f,z)$,
we have the following result.

\bthml{finalbd}
The size of the subdivision tree of the \texttt{Soft Root Clustering} algorithm is bounded by
        \beql{finalbd}
        O\paren{|S_0| + \int_{B_0 \sm \su_{\calc \in S_0}D_\calc/3} \gamma_f(z)^2 \da}
        = O\paren{|S_0| + \int_{B_0 \sm \su_{\calc \in S_0}D_\calc/3} \frac{\da}{R_f(z)^2}},
        \eeql
where $\da$ denotes the area form.
\ethml

We will illustrate with some examples  later that the two bounds do not always yield the same quantity,
in fact, working with $\gamma(f,z)$ seems to give better bounds.
However, there are occasions when the geometric bound is easier to handle than the analytic one.

\subsection{Number of exclusion boxes in a special case}
\label{sec:numb-excl-boxes}

There is a special extreme case where $S_0$ is empty and where the bound above may not be 
satisfying: The case when $D(m_\calc, r_\calc) \si 4B_0 = \es$ and $B_0~\ib~D_\calc$.
That means $B_0$ is well inside $D_\calc$ but it is so small that no box in the subdivision tree may
detect $\calc$, because the  condition $D(m_\calc, r_\calc) \si 4B_0$ ensures that 
for all points $p \in B_0$ we have $|p -m_\calc| + r_\calc > 4r_{B_0}$, and so no
box in a subdivision of $B_0$ will satisfy the converse for inclusion in  \refLem{geomconv};
for an  illustration see \refFig{special}.
What is the number of boxes
in the subdivision tree required to detect exclusion so close to a cluster? 
We derive an independent bound on $\int_{B_0} \gamma(f,z)^2 \da$ in this special case.

\vfigpdf{Illustration of the special case}{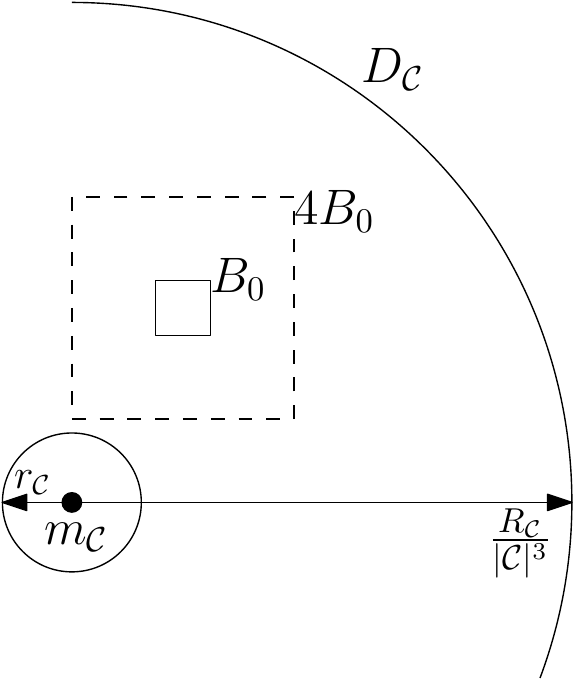}{0.75}

Recall the definition of $S_g(z)$ in \refeq{sgz}. From \refeq{gammaub} we know that
        $$\gamma(f,z)^2 \le 2 (S_g(z)^2 + \gamma_\calc(z)^2).$$
Therefore, 
        $$\int_{B_0} \gamma(f,z)^2 = O\paren{\int_{B_0} S_g(z)^2 \da + \int_{B_0} \gamma_\calc(z)^2 \da}.$$
Since by assumption $D(m_\calc, r_\calc)$ does not intersect $4B_0$, it follows that
for all $z \in B_0$, and $\alpha \in \calc$, $|z-\alpha| \ge r_{B_0}$. Therefore,
$S_g(z) \le |\calc|/r_{B_0}$. For all 
$z \in B_0 \ib D_\calc$, it also follows that  $\gamma_\calc(z) \le 4 \gamma_\calc(m_\calc)$. 
Therefore, the number of exclusion boxes is bounded by
        $$O\paren{\int_{B_0} \frac{|\calc|^2}{r_{B_0}^2} \da + \gamma_\calc(m_\calc)^2 w(B_0)^2}.$$
The assumption $B_0 \ib D_\calc$ implies that $w(B_0) \gamma_\calc(m_\calc) < 1$, 
and hence the second quantity in the right-hand side above can be bounded by a constant.
The area term in the first quantity cancels out with the denominator, 
to conclude that the total number of exclusion boxes is  at most $O(|\calc|^2)$. 
What is the width of the smallest box? Since there are at most $O(|\calc|^2)$ boxes, the depth of the
subdivision tree is $O(\log |\calc|)$, and therefore the width of the smallest box is $\Omega(r_{B_0}/|\calc|)$.
More formally, at every parent of a leaf box $B$, the exclusion predicate failed: therefore, 
$2c_3 \gamma(f,z) r_B \ge 1$. Again, by the upper bound from \refeq{gammaub} we have
        $$2c_3 (S_g(z) + \gamma_\calc(z)) r_B \ge 1.$$
Substituting the upper bound $S_g(z) \le |\calc|/r_{B_0}$ and $\gamma_\calc(z)\le 4 \gamma_\calc(m_\calc)$, we obtain that
        $$2 c_3\paren{\frac{|\calc|}{r_{B_0}} + \gamma_\calc(m_\calc)}r_B \ge 1.$$
Using the inequality $w(B_0) \gamma_\calc(m_\calc) < 1$, we further get that
        $$r_B= \Omega \paren{\frac{r_{B_0}}{|\calc|}},$$
as desired

\subsection{Number of exclusion boxes for the exponential function}
\label{sec:exp}
We now derive an explicit form for the bound given in \refThm{finalbd}
on the number of subdivisions for the exponential function. For the sake of understanding,
we will derive bounds on both the integrals given in the theorem.

Let the input box $B_0$ be centered at the origin, and its width
be a dyadic number.
Clearly, $S_0$ is empty as there are no roots. It is not hard to see that
$\gamma_{\exp}(z) = 1$ and so the first integral in \refThm{finalbd} is $O(\text{Area}(B_0))$.

To bound the second integral, we need to derive a lower bound on $R_{\exp}(\alpha)$.
The radius of univalence of $\exp(z)$ is less than $ 2\pi$, due to its periodicity, however, 
we are interested in the
radius of univalence of $(z-\alpha)e^z$. From the lower bound in \refeq{roulbd}, we have that
\begin{align*}
        R_{\exp}(\alpha) &\ge |e^\alpha| \sup_r \frac{r}{\sup_{z\in D(\alpha,r)} |e^z||1+(z-\alpha)|}\\
                            &=  \sup_r \frac{r}{\sup_{z\in D(\alpha,r)} |e^{(z-\alpha)}||1+(z-\alpha)|}.  
\end{align*}
Observe that the supremum of the denominator is attained when $(z-\alpha)=r$, in which case,
from basic calculus, we obtain that the supremum of the right-hand side is $\phi/(e^\phi(1+\phi))$, where
$\phi := (\sqrt{5}-1)/2$. Thus the integral 
        $$\int_{B_0} \frac{\da}{R_{\exp}(z)^2} = O(\text{Area}(B_0)).$$

Both the integral bounds in \refThm{finalbd} give us the same result. 
The bound, however, is exponential in the bit-representation of the coordinates of the endpoints of $B_0$. 
Is the bound tight?
Intuitively, yes since when $e^m$ dominates $r e^z$, where $z$ is a boundary point of $D(m,r)$, we
would have $r$ to be smaller than one, i.e., when the exclusion predicate holds on
a subdivision of $B_0$, all the boxes in the subdivision will have width below one. 
More precisely, if the exclusion test holds 
then  $|e^m| \ge r \sup_{z \in \partial D(m,r)}|e^z|$; note that we have replaced
the box-function of the derivative with the exact range. This is equivalent to
$r \le \inf_{z \in \partial D(m,r)}|e^{m-z}| = e^{-r}$ , which implies that $.5 < r < 1$, therefore, corroborating
our intuition.

\subsection{Number of exclusion boxes for the sine function} \label{sec:sine}
We know that the roots of the sine function are $k\pi$, where $k \in \ZZ$.
Let the input box $B_0$ be in the upper half plane 
such that $6B_0$ does not intersect the real axis and
$\Im z \ge 1/2$, for all $z \in B_0$. This ensures that $S_0$ is again empty.

We first derive tight estimates on $|\sin z|$ for $z \in B_0$. Let 
the real part be $\Re(z)=x$ and the imaginary part be $\Im(z)=y$.
From the definition of the sine function, we know that
        $$\sin z = \frac{e^{ix- y} - e^{-ix + y}}{2i} = \frac{e^{y-ix}}{2i} \paren{-1+e^{2(ix-y)}}.$$
Taking absolute values on both sides and applying the triangular inequality we get that
        $$|\sin z| \le \frac{e^y}{2}\paren{1 + e^{-2y}} < \frac{3}{4}e^{\Im(z)},$$
where in the last inequality we use the fact that $2y \ge 1$. The same argument also shows
that 
        $$|\sin z| > \frac{e^{\Im(z)}}{4}.$$
Similarly, one can show that for all $z \in B_0$
        $$\frac{e^{\Im(z)}}{4}< |\cos z| < \frac{3}{4}e^{\Im(z)}.$$

From these bounds it follows that $\gamma(\sin,z) = O(1)$ and hence the first integral
bound in \refThm{finalbd} on the size of the subdivision tree is $O(\text{Area}(B_0))$.
Substituting the bounds on $|\sin z|$ above in \refeq{roulbd}, we obtain that
        $$R_{\sin}(\alpha) \ge \sup_r \frac{r}{3 |e^{\Im (z-\alpha)}|\cdot|1+(z-\alpha)|}.$$
The supremum is attained when $\Im(z-\alpha) = r$, and it is $\phi/(3e^\phi (1+\phi))$.
Therefore, $R_{\sin}(\alpha) = \Omega(1)$, for all $\alpha \in B_0$ and hence
the bound in \refThm{finalbd} on the number of subdivisions required to exclude $B_0$ 
is $O(\text{Area}(B_0))$. Hence, we see that both the integral bounds in \refThm{finalbd}
yield an exponential bound on the size of the subdivision tree in the input representation.
We next argue that this is tight.

One can again show that when the exclusion test holds, the boxes have width smaller than one.
More precisely, suppose $\sin m \neq 0$, and suppose
        $$\abs{\sin m} \ge r \sup_{z \in \partial D(m,r)} \abs{\cos z};$$
we have again replaced the box-function for the derivative by its exact range.
From the upper bound on $|\sin m|$ and the lower bound on $|\cos z|$ we know that the inequality
above implies that
        $$3 e^{\Im(m)} > r \sup_{z \in \partial D(m,r)} e^{\Im(z)};$$
Since $z$ is on the boundary of $D(m,r)$, the right-hand side above is maximized when $z = m + ir$, i.e., 
$\Im(z) = \Im(m) + r$.
Therefore, when the exclusion test is successful, we will have $3 \ge re^r$, which implies that $r < 2$.
This means that $B_0$  will be subdivided into a uniform grid of boxes
with width smaller than some constant before the exclusion test is successful. Hence the integral bounds
derived above are tight.

\subsection{The case of polynomials}\label{sec:poly}
Let $f$ be a monic polynomial of degree $d$ and $Z(f)\ibp \CC$ be its multiset of roots. 
Let $\Mea(f) \as \prod_{\alpha \in Z(f): |\alpha| > 1} |\alpha|$ be the \dt{Mahler measure} of $f$; 
it follows from the definition that $\Mea(f)$ is an upper bound on the absolute value of all the roots
of the polynomial; moreover, from
Landau's inequality \cite[p.~390]{rahman-schmeisser:bk} we know that for a monic $f$, $\Mea(f) \leq  \sqrt{1 + \sum_{i=0}^{d-1} |f_i(0)|^{2}}$.
We will derive explicit upper bounds for the two integral
bounds given in \refThm{finalbd}.
We start with a more geometric understanding of strongly separated clusters in the polynomial setting.

Let $\calc$ be a cluster of roots of $f$, $m_\calc$ its centroid, and $r_\calc$ the radius of the smallest disc
centered at $m_\calc$ containing all the points in $\calc$. Additionally define 
        \beql{sigmac}
        \sigma_\calc \as \min_{\alpha \in Z(f)\sm \calc}|m_\calc - \alpha|,
        \eeql
i.e., the distance from the centroid
$m_\calc$ to the nearest root of $f$ outside the cluster $\calc$; it is infinite when $\calc$
is the set of all roots $Z(f)$. Formally speaking, this is not the same $R_\calc$ as defined in~\refeq{Rcalc},
however, as it will soon become clear it plays an analogous role.
The function underlying the definition of the $\gamma_\calc$-function is
        $$h(z) = \prod_{\alpha \in Z(f)\sm \calc} (z-\alpha),$$
and hence
        $$\gamma_\calc(m_\calc) = \sup_{k \ge 1}\abs{\frac{h_k(m_\calc)}{h(m_\calc)}}^{1/k}
                                                       \le \sum_{\alpha \in Z(f) \sm \calc} \frac{1}{|m_\calc-\alpha|}
                                                       \le \frac{(d-|\calc|)}{\sigma_\calc}.$$
This implies that
        $$R_\calc = \frac{|\calc|}{c_0 \gamma_\calc(m_\calc)} \ge \frac{|\calc|}{c_0 (d-|\calc|)} \sigma_\calc.$$
Hence the disc $D_\calc$, defined in \refeq{dc}, 
has radius
        \beql{radineq}
        \frac{R_\calc}{|\calc|^3} \ge \frac{\sigma_\calc}{\paren{c_0 |\calc|^2 (d-|\calc|)}} \ge \frac{4\sigma_\calc}{(c_0 d^3)}, 
        \eeql
where in the last step we use the inequality that the geometric mean of two quantities
is smaller than its arithmetic mean and $|\calc| \le d$. 
When $\calc$ is a multiple root, say $\alpha$, of $f$ then $\sigma_\alpha$ denotes the separation
between $\alpha$ and a nearest distinct root of $f$, and $m_\alpha$ denotes its multiplicity.


To derive explicit forms for the two bounds given in \refThm{finalbd},  the key result is an
upper bound on the number of exclusion boxes in both bounds, as that turns out to be
the dominating term. 
We will first derive an upper bound on the following integral:
        \beql{ib0}
        \int_{B_0 \sm \su_{\calc \in S_0} D_\calc/3} \frac{\da}{R_f(z)^2},
        \eeql
where $S_0$ is as defined in \refeq{s0}. 
To bound this integral, we need a lower bound on $R_f(\beta)$, 
where $\beta \nin Z(f)$. For a polynomial $g \in \CC[x]$, let 
$d(\beta, Z(g))$ denote the distance from $\beta$ to the nearest root of $g$.
From the theorem of Alexander-Kakeya \cite[Thm.~4.3.2]{rahman-schmeisser:bk}, we have the following:
\bpropl{ak}
For $\beta \nin Z(f)$, define $g \as (z-\beta)f$. Then
        $$R_f(\beta) \ge d(\beta, Z(g')) \sin(\pi/(d+1)).$$ 
\epropl


For $z\in \CC$, define 
        \beql{sfz}
        S_f(z) \as \sum_{\alpha \in Z(f)} \frac{1}{|z-\alpha|}.
        \eeql
We claim the following:
\blem
Let $0 < \eps< 1$, $\beta \nin Z(f)$ and $g\as (z-\beta)f$. Then
        $$d(\beta, Z(g')) \cdot S_f(\beta) \ge \max \set{\eps, 1-\eps}.$$ 
\elem
\bpf
Let $\tau \in Z(g')$ be a nearest critical point of $g$ to $\beta$; note that $\beta$ is a simple root of $g$
and hence $\tau$  is not equal to $\beta$. Our first claim is that
        $$S_f(\tau) |\beta - \tau| \ge 1.$$
The second claim is that if $S_f(\beta)|\beta -\tau| \le \eps$ then
        $$S_f(\tau) (1-\eps) \le S_f(\beta)$$
and hence combined with the first claim we get the desired inequality. 

The first claim is straightforward: Since $\tau$ is a critical point of $(z-\beta)f$, we know that
        $$(\beta - \tau) f'(\tau) + f(\tau) = 0.$$
Therefore, 
        $$|\beta - \tau| \abs{\frac{f'(\tau)}{f(\tau)}} = 1$$
and hence by the triangular inequality
        $$S_f(\tau) |\beta - \tau| \ge 1.$$

If $S_f(\beta) |\beta - \tau| \le \eps$ then it follows that $|\beta - \tau| \le \eps |\beta - \alpha|$
for all $\alpha \in Z(f)$. So
        $$|\tau - \alpha| \ge |\beta - \alpha| - |\beta - \tau | \ge |\beta - \alpha| (1-\eps).$$
Dividing both sides of the inequality above with $|\beta-\alpha|\cdot |\tau-\alpha|$ 
and adding over all $\alpha$ we obtain that
        $$S_f(\tau) (1-\eps) \le S_f(\beta).$$

\epf

Taking $\eps = 1/2$ in the lemma above along with the lower bound in \refPro{ak},
we obtain that
        \beql{Rfb}
        R_f(\beta)= \Omega\paren{\frac{1}{2dS_f(\beta)}}.
        \eeql
Applying this inequality in \refeq{ib0}, we get the following upper bound on the 
number of subdivisions for the exclusion test:
        \beql{ib11}
        O\paren{d^2\int_{B_0 \sm \su_{\calc \in S_0} D_\calc/3} S_f(z)^2 \da}.
        \eeql
Applying the Cauchy-Schwarz inequality to the quantity $S_f(z)^2$ we get that
        $$S_f(z)^2 \le d \sum_{\alpha \in Z(f)} \frac{1}{|z-\alpha|^2}.$$
Substituting this in \refeq{ib11} and switching the order of summation and the integral
we get the following upper bound on the number of subdivisions:
        \beql{ib12}
        O\paren{d^3\sum_{\alpha \in Z(f)} \int_{B_0 \sm\su_{\calc \in S_0} D_\calc/3} \frac{\da}{|z-\alpha|^2}}.
        \eeql
Every root $\alpha \in Z(f)$ belongs to one of two types of clusters: it is either in a cluster
in $S_0$ or in the set $S_1$ defined as
        \beql{s1}
        S_1 \as \set{\text{maximal strongly separated clusters $\calc$ partitioning roots of $Z(f)$ outside $2B_0$}}.
        \eeql
Therefore, the summation
in the integral above can be equivalently expressed as the sum over all clusters $\calc$ in $S_i$, and
then over all roots in $\calc$, where $i=0, 1$. The sum corresponding to $S_0$ is 
        \beql{ib14}
        O\paren{d^3\sum_{\calc \in S_0}\sum_{\alpha \in \calc }
                 \int_{B_0 \sm \su_{\calc' \in S_0}D_{\calc'}/3} \frac{\da}{|z-\alpha|^2}}.
        \eeql
Observe that the set 
        $$B_0 \sm \su_{\calc' \in S_0} D_{\calc'}/3 \ib B_0 \sm D_{\calc}/3$$
for a given $\calc \in S_0$, which implies that the integral above is upper bounded by
        \beql{ib15}
        O\paren{d^3\sum_{\calc \in S_0}\sum_{\alpha \in \calc }
                 \int_{B_0 \sm D_{\calc}/3} \frac{\da}{|z-\alpha|^2}}.
        \eeql
Given a cluster $\calc$,  for all $\alpha \in \calc$ and $z$ outside of 
$D_\calc/3$, we have
        $$|z-\alpha| \ge |z-m_\calc| - |m_\calc-\alpha| \ge |z-m_\calc| - r_\calc \ge \frac{|z-m_\calc|}{2},$$
where in the last inequality we have used the definition of a strongly separated cluster from \refeq{ssc}.
Substituting this in \refeq{ib15}, we obtain the following upper bound
        \beql{ib16}
        O\paren{d^3\sum_{\calc \in S_0}|\calc|
                 \int_{B_0 \sm \su_{\calc \in S_0}D_\calc/3} \frac{\da}{|z-m_\calc|^2}}.
        \eeql
Extending the integral for each $\calc$ over the annulus centered at $m_{\calc}$ 
with inner radius $R_{\calc}/(3|\calc|^3)$ and outer radius $2r_{B_0}$, 
we obtain that the integral term in \refeq{ib16} is bounded by
        $$\int_{R_{\calc}/(3|\calc|^3)}^{2r_{B_0}} \int_{0}^{2\pi} \frac{r dr d\theta}{r^2} 
                = O\paren{\log \frac{w(B_0)|\calc|^3}{R_{\calc}}},$$
and so the integral in \refeq{ib16}, is bounded by
        \beql{ib17}
        O\paren{d^3
          \paren{\sum_{\calc \in S_0}\paren{|\calc| \log \frac{w(B_0)|\calc|^3}{R_{\calc}}}}}.
        \eeql

Now consider the sum analogous to \refeq{ib14} for the set of clusters in $S_1$:
        \beql{ib24}
        O\paren{d^3\sum_{\calc \in S_1}\sum_{\alpha \in \calc }
                 \int_{B_0 \sm \su_{\calc' \in S_0}D_{\calc'}/3} \frac{\da}{|z-\alpha|^2}}.
        \eeql
Since $S_1$ contains roots outside $2B_0$, it follows that for all $\calc \in S_1$, $m_\calc$ is not in $2B_0$.
Therefore, we can take the integral over all of $B_0$ to obtain the following bound
        \beql{ib25a}
        O\paren{d^3\sum_{\calc \in S_1}\sum_{\alpha \in \calc }
                 \int_{B_0} \frac{\da}{|z-\alpha|^2}}.
        \eeql
Again, for all $\alpha \in \calc$, $|z-\alpha| \ge |m_\calc - z|/2$. Therefore, we obtain
        \beql{ib25}
        O\paren{d^3\sum_{\calc \in S_1}|\calc|\int_{B_0} \frac{\da}{|z-m_\calc|^2}}.
        \eeql
Switching to polar coordinates, as was done earlier, and taking the integral 
on the annulus with inner radius the distance of $m_\calc$ to the box $B_0$, which is at least
$w(B_0)$, and outer radius $2r_{B_0}$, we get the following bound
        \beql{ib26}
        O\paren{d^3\sum_{\calc \in S_1}|\calc|}  = O(d^4).
        \eeql
Adding this bound to the bound in \refeq{ib17}, and substituting the lower bound on
$R_\calc/|\calc|^3$ from \refeq{radineq}, we obtain the following upper bound on 
\refeq{ib0}, which is the second integral in \refThm{finalbd}:
        \beql{ib27}
        O\paren{d^4 \log w(B_0) + d^4 \log d - d^3 \sum_{\calc \in S_0}|\calc| \log \sigma_\calc}.
        \eeql
If we consider the integral 
        $$\int_{B_0 \sm \su_{\calc \in S_0} D_\calc/3} \gamma(f,z)^2 \da$$
in \refThm{finalbd}, use the upper bound $\gamma(f,z) \le S_f(z)$, as shown in \refeq{gammaub} 
where $h$ is taken to be the constant polynomial, and follow
the previous argument from \refeq{ib12} onwards, then the bound in \refeq{ib27} 
improves by a factor quadratic in the degree: 
\bthml{polybd}
The size of the subdivision tree of the \texttt{Soft Root Clustering} algorithm for a polynomial
$f$ on a box $B_0$ is bounded by
        $$O\paren{d^2 \log w(B_0) + d^2 \log d - d \sum_{\calc \in S_0}|\calc| \log \sigma_\calc}.$$
\ethml

If we assume that $f$ is an integer polynomial, and the initial box $B_0$ is centered
at the origin, contains all the roots of $f$
and has radius $\Mea(f)$ then the bound can be made more explicit.
\bcorl{intpoly}
The  size of the subdivision tree of the \texttt{Soft Root Clustering} algorithm for an
{\em integer} polynomial $f$ is bounded by $O(d^2 \log \Mea(f) + d^3)$.
\ecorl
\bpf
Taking $S_0$ to be comprised of the roots $\alpha$ along with their multiplicity $m_\alpha$,
the summation term  in \refThm{polybd} is 
        $(-\sum_{\alpha \in Z(f)}m_\alpha \log \sigma_\alpha).$
From \cite[Thm.~2]{emeliyanenko-sagraloff:2012}, this is at most $O(d^2 + d\log \Mea(f))$,
which yields the desired bound.
\epf

The estimate of Theorem 1.1 in \cite{yakoubsohn:bisection-analysis:05}
(also see Section 9.1 of the paper) bounds the number of subdivisions 
by $O(d^3 \log (w(B_0)\gamma(f)))$, for a square-free polynomial $f$ of degree $d$, where $\gamma(f):=\max_{z_i:f(z_i)=0}\gamma(f,z_i)$. 
It can be shown that $\gamma(f) = O(1/\sigma(f))$, where $\sigma(f) \as \min_\alpha \sigma_\alpha$,
i.e., the worst case root separation bound. From well known root separation bounds
\cite[Chap.~6]{yap:bk}, we know that $(-\log \sigma_\alpha) = O(d(\log \Mea(f)+\log d))$.
Therefore, the bound in \cite{yakoubsohn:bisection-analysis:05} is of the order 
        $$O(d^4(\log \Mea(f)+\log d)),$$
where $B_0$ is again the box centered at the origin with radius $\Mea(f)$.
It follows that the result in \refCor{intpoly}  is at least an improvement by  a linear factor (in the degree) over the result in 
\cite{yakoubsohn:bisection-analysis:05}, and without the assumption that $f$ is square-free.

\section{Bounding the Precision}\label{sec:precision}
In this section, we derive a bound on the precision requested by the algorithm from the
oracles for the function and its higher order derivatives.

In the procedure \texttt{firstC}, we invoke the predicate 
$\intbox \wtC_k(s B)$, for $k=0 \dd N_0$, some scaling factor $s$, and
a box $B \ib B_0$ produced in the subdivision. 
Since the box $B$ is not necessarily a terminating box, it is possible that we check
all the predicates $\intbox \wtC_k(sB)$, for $k=0 \dd N_0$. For a given $k$,
the procedure \texttt{SoftCompare} compares the following two quantities:
        \beql{akbk}
        \calA_k \as |f_k(m_B)| (sr_B)^k \text{ and } \calB_k \as \sum_{i=0}^{k-1}|f_i(m_B)|(sr_B)^i + |\intbox f_{k+1}(sB)|(sr_B)^{k+1}.
        \eeql
From \refLem{softcompare}, we know that the precision requested from the corresponding
oracles for a given box $B$ is upper bounded by 
        $$
        O\paren{- \log\omin\max \set{\calA_k,  \calB_k }},
        $$
where the $\omin$ function is defined as in \refeq{omin}.
Taking the minimum over all boxes $B$ in the subdivision tree, we obtain the following
bound on the precision requested from the oracles:
        \beql{precbd}
        O\paren{- \log \omin_{B}\paren{\max \set{\calA_k,  \calB_k }}}.
        \eeql
Therefore, we need to derive a lower bound on $\max \set{\calA_k, \calB_k}$, for $k=0 \dd N_0$, 
anf for all boxes $B$ produced in the subdivision. This  entails  a 
lower bound on the width of the inclusion and exclusion boxes produced in the subdivision.
In light of \refLem{widthinc}, we assume that the subdivision
continues until the boxes that detect a cluster are exactly those that satisfy the conditions of the lemma,
i.e., a box is an inclusion box if and only if  it satisfies the condition of the lemma.
The resulting tree is a further subdivision of the tree $T_0$ defined in \refSec{integral-bound},
and since the boxes only shrink, it suffices to derive a lower bound on their width, which is what we do next.

\subsection{Lower Bound on Width of Inclusion and Exclusion Boxes}
\label{sec:lower-bound-width}
Since by our assumption all inclusion boxes $B$ satisfy the condition of \refLem{widthinc},
from the proof of the lemma it follows that their width is at least
        \beql{incbd}
        \min_{\calc \in S_0} \frac{1}{6c_0 \gamma_\calc(m_\calc) |\calc|^2}.
        \eeql
Clearly, $|\calc| \le N_0$, since $N_0$ counts the number of roots of $f$ in $B_0$
with multiplicity.
Also, note that for all $\calc \in S_0$, we have from \refeq{assum1} that
        $$r_{B_0} \ge R_\calc/|\calc|^3 \ge 8 r_\calc,$$ 
where the last inequality holds because $\calc$ is a strongly separated cluster.
Therefore, $r_{B_0} \ge 2r_\calc$, for all $\calc \in S_0$, and we can
apply \refeq{gammacbound} with $r  = r_{B_0}$.
Since $\calc \in S_0$, we have  $m_\calc\in 2B_0$, and hence 
$D(m_\calc, 2r_{B_0}) \ib 4B_0$. Therefore, we can express the bound in \refeq{gammacbound} 
for $r=r_{B_0}$ as
        $$\frac{1}{\gamma_\calc(m_\calc)} = \Omega\paren{\frac{r_{B_0}^{|\calc|+1} \abs{f_{|\calc|}(m_\calc)}}{2^{|\calc|}M(f, 4B_0)}}.$$
Substituting this in \refeq{incbd}, we get the following:
\bleml{inclb}
The width of all inclusion boxes is at least
        $$\Omega \paren{\frac{\omin\paren{r_{B_0}}^{N_0+1}}{2^{N_0}M(f, 4B_0) N_0^2}\cdot  \omin_{\calc \in S_0} \abs{f_{|\calc|}(m_\calc)}}.$$
\eleml

The argument in \refSec{numb-excl-boxes} handles the exclusion boxes in the special case.
What remains is a lower bound on the exclusion boxes in general.
From \refLem{widthinc} we deduce that there will be a constant number of inclusion boxes covering
$D_\calc/3$, for a cluster $\calc \in S_0$. Therefore, if $B$ is an exclusion box then
 it follows that the $m_B \nin D_\calc/3$ for all strongly separated clusters $\calc\in S_0$.
At the parent $B'$ of $B$, we know from \refLem{excl} 
that for all $z\in B'$, $2c_3 \gamma(f,z) r_B \ge 1$.
In particular, $2 c_3 \gamma(f, m_B) r_B \ge 1$, which implies
        $$r_B = \Omega(1/\gamma(f, m_B)).$$
Incidentally, this is also tight: since the exclusion test worked for $B$, it follows that Pellet's condition must hold
        $$|f(m_B)| \ge \sum_{k > 0} |f_k(m_B)| r^k$$
which instead implies that $r < \inf |f(m)/f_k(m)|^{1/k} = 1/\gamma(f,m_B)$.
How large is $\gamma(f, m_B)$, given that $|m_B - m_\calc| \ge R_\calc/(3|\calc|^3)$, for all strongly
separated clusters $\calc \in S_0$? From \refeq{gammafbd}, it follows that 
        $$\gamma(f,m_B) = O(M(f,m_B,2r)/(r|f(m_B)|),$$
for all $r \ge 0$. Choosing $r \as r_{B_0}$, and observing that $D(m_B, 2r_{B_0}) \in 3B_0$ we
get that $\gamma(f,m_B) = O(M(f, 3B_0)/(r_{B_0}|f(m_B)|))$, and hence the width of an exclusion box $B$ is
$\Omega(|f(m_B)|r_{B_0}/M(f, 4B_0)|)$.  The crucial part is the quantity $|f(m_B)|$.
How small can this get? If the set $S_0$ is empty then we cannot rely on clusters to bound this quantity
away from zero. Therefore, the following quantity
        \beql{minpar}
        m(f, B_0) \as \inf\set{|f(z)|: z \in B_0 \sm \su_{\calc \in S_0}D_\calc/3}
        \eeql
is a natural parameter for measuring the complexity. Thus we obtain the following:
\bleml{exclb}
The width of all exclusion boxes is at least 
        $$\Omega\paren{\frac{r_{B_0} m(f,B_0)}{N_0M(f, 4B_0)}}.$$
\eleml
The quantity $N_0$ in the denominator is to accommodate the result of \refSec{numb-excl-boxes}.

\refLem{inclb} and \refLem{exclb} can be combined by replacing $r_{B_0}$ in \refLem{exclb}
with $\omin(r_{B_0})$ and taking the product of the two bounds to give the following:
\bleml{widthbd}
The width of any box in the subdivision tree is at least
        $$\Omega \paren{\frac{\omin \paren{r_{B_0}}^{N_0+2}m(f,B_0)}{2^{N_0}M(f, 4B_0)^2 N_0^3}
                \cdot  {\omin_{\calc \in S_0}\paren{\abs{f_{|\calc|}(m_\calc)}}}},$$

\eleml

We remark that with a more careful argument the constant $4$ in $M(f, 4B_0)$ can be improved to $(2+\eps)$,
for  fixed $\eps > 0$.
\subsection{Lower bound on higher order derivatives}
\label{sec:lower-bound-higher}
We now derive a lower bound on $\max_k \set{\calA_k, \calB_k}$, where the two quantities
are defined in \refeq{akbk}. Why is this quantity bounded away from zero?
Consider $\max_k |f_k(m)|$, for all $k$ and $m \in B_0$. 
If $m$ is not a root of $f$, then at least one of $\calA_k$ or $\calB_k$ is at least $|f(m)|$.
Otherwise, not  all the $N_0$ derivatives can vanish at $m$ as $N_0$ is an upper bound on the number of roots in
$B_0$; in the worst case $m$ is a root with multiplicity $N_0$ but even in that case the
$N_0$th derivative is non-zero.

Consider a box $B$ in the subdivision tree. There are two cases to consider:
\begin{enumerate}[{Case} 1.]
\item Either $m_B$ is in the disc $D_\calc/3$, for some strongly separated cluster $\calc \in S_0$, or
\item $m_B$ is  not in any of the discs $D_\calc/3$, for all $\calc \in S_0$.
\end{enumerate}

Let us begin with the latter case. In this case, we know that $|f(m_B)| \ge m(f, B_0)$ (see \refeq{minpar}). 
For $k=0$,
it is clear that $\calA_0 = |f(m_B)| > m(f,B_0)$. For $k >0$, the term $\calB_k$ always includes the term
$|f(m_B)|$ in the summation, therefore, $\calB_k \ge |f(m_B)|$, and hence, 
        \beql{case1}
        \max_k \set{\calA_k, \calB_k} \ge m(f, B_0),
        \eeql
in this case.

Now we consider the first case, i.e., when $m_B\in D_\calc/3$, for a strongly separated cluster $\calc$ 
of size $j \as |\calc|$.
In this case, we know from \refLem{fkzbd} that $|f_{j}(m_B)| > |f_{j}(m_\calc)|/2$. 
The argument for Case 2 can be generalized to work in this case.
We want to derive a lower bound on $\max \set{\calA_k, \calB_k}$.
However, there are three cases to consider depending on how $k$ compares with $j$:
\begin{itemize}
\item When $k=j$, $\calA_j = |f_j(m_B)|(sr_B)^j  = \Omega(|f_j(m_\calc)|(sr_B)^j)$.

\item If $k > j$, then $\calB_k$ includes the absolute value $|f_j(m_B)|(sr)^j$ in the summation, therefore,
  $\calB_k > \calA_j$.
\item If $k < j$ then 
        $$\calB_k \ge |\intbox f_{k+1}(D(m_B,sr_B))|(sr_B)^{k+1} \ge M(f, k+1, m_B, sr_B) (sr_B)^{k+1}.$$ 

   Therefore, 
        $$\calB_k \ge (sr_B)^{j+1} |f_j(m_B)|=\Omega(|f_j(m_\calc)| (sr_B)^{j+1}).$$
\end{itemize}
To summarize, when $m_B \in D_\calc/3$  for a strongly separated cluster $\calc \in S_0$ then
        \beql{case2}
        \max_k \set{\calA_k, \calB_k} = \Omega(|f_{|\calc|}(m_\calc)| (sr_B)^{|\calc|}).
        \eeql

Taking the minimum over all $B$ in both \refeq{case1} and \refeq{case2}, 
along with the observation that $N_0$ 
dominates $|\calc|$ for all $\calc \in S_0$, we obtain
        $$\min_B \max_k \set{\calA_k, \calB_k} 
        = \Omega\paren{\omin_{\calc\in S_0}\paren{|f_{|\calc|}(m_\calc|)} \cdot \omin\paren{r_B}^{N_0}
                \cdot \omin\paren{m(f,B_0)}}.$$
Substituting this in \refeq{precbd}, and using the lower bound on $(\min_B r_B)$ from 
\refLem{widthbd} we obtain the following result after some manipulations:

\bthml{precisionb}
The bit-precision requested from the oracles by \texttt{SoftCompare} is bounded by 
        $$O\paren{N_0\log \frac{M(f, 4B_0)}{m(f, B_0)} + N_0^2 - N_0^2 \log \omin(r_{B_0}) 
                -N_0(\log \omin_{{\calc \in S_0}} \abs{f_{|\calc|}(m_\calc)})}.$$
\ethml


\subsection{Bit-complexity for polynomials}
\label{sec:bit-compl-polyn}
In this section, we specialize the result above for the case of polynomials
in order to state the bound in terms of more natural parameters analogous to \refThm{polybd},
and get a result similar to \cite[Thm.~A]{becker+4:cluster:16}.
There are only two terms that demand our attention: $M(f, 4B_0)$  and $m(f, B_0)$. 
We need an upper bound on the former, and a lower bound on the latter.

The upper bound is relatively straightforward. Let $\tau \in \RR_{\ge 1}$ be such that
the absolute value of the coefficients of $f$ and any point in $B_0$ is upper bounded by $2^{\tau}$.
Then it is straightforward to show that 
        $$M(f, 4B_0) \le (d+1) 2^{O(d\tau)},$$ 
which implies
        $$\log M(f, 4B_0) = O(d\tau).$$
Recall that $\Mea(f)$ denotes the Mahler measure of $f$. Using the fact that the
coefficients are elementary symmetric polynomials in the roots, and 
assuming that $B_0$ is a box contained in the disc  of radius $\Mea(f)$ centered at the origin, 
we obtain  that
        $$\tau = O(d \log \Mea(f)),$$
whence
        \beql{logM}
        \log M(f, 4B_0) = O(d^2 \log \Mea(f)).
        \eeql

We now derive a lower bound on $m(f, B_0)$. Let $z \in B_0 \sm \su_{\calc \in S_0} D_\calc/3$.
Our aim is to derive a lower bound on $|f(z)|$, which will follow if we derive a lower bound 
on $|z-\alpha|$, where $\alpha$ is a root of $f$, and take the product over all roots of $f$. 
There are two cases to consider:
\begin{enumerate}
\item If $\alpha \in \calc$ for some $\calc \in S_0,$ then it is true that
        $$|z-\alpha| \ge |z-m_\calc| - |m_\calc - \alpha|\ge |z-m_\calc| - r_\calc \ge \frac{R_\calc}{2|\calc|^3},$$
 since $\calc$ is a strongly separated cluster, and thus $r_\calc \le R_\calc/(8|\calc|^3)$.
 From \refeq{radineq}, it follows that
        $$|z-\alpha| \ge \frac{2\sigma_\calc}{c_0d^3}.$$
 
\item If $\alpha$ is a root outside $2B_0$, then since $z \in B_0$ it trivially follows that
        $$|z-\alpha| \ge r_{B_0}.$$
\end{enumerate}
From these two inequalities we obtain that
        \begin{align*}
          |f(z)| &= \paren{\prod_{\calc \in S_0} \prod_{\alpha \in \calc} |z- \alpha|}\paren{\prod_{\alpha \nin 2B_0}|z-\alpha|}\\
                    &\ge \paren{\prod_{\calc \in S_0}\paren{\frac{2\sigma_\calc}{c_0d^3}}^{|\calc|} } \cdot \paren{\omin(r_{B_0})^d}.
        \end{align*}
Therefore, 
        \beql{logm}
        -\log m(f, B_0) = O\paren{-\sum_{\calc \in S_0} |\calc| \log \sigma_\calc + d \log d - d \log \omin(r_{B_0})}.
        \eeql
Substituting \refeq{logM} and the bound above in \refThm{precisionb}, we obtain the following:
\bcorl{precisionpoly}
Let $f$ be a degree $d$ polynomial. The bit-precision requested from the oracles by \texttt{Soft-Compare}
is bounded by
        $$O\paren{d^3 \log \Mea(f) - d^2 \log \omin(r_{B_0}) - d\sum_{\calc \in S_0} |\calc| \log \sigma_\calc
          -d(\log \omin_{{\calc \in S_0}} \abs{f_{|\calc|}(m_\calc)})}.$$
\ecorl
This bound is similar to the one given in \cite[Thm.~A]{becker+4:cluster:16}, especially, in the appearance
of the summation term above. The result can be further specialized to the situation where 
$f$ is an integer polynomial, and $B_0$ is centered at the origin and has radius $\Mea(f)$, as was done in 
\refCor{intpoly}.

\section{Remarks}
\label{sec:remarks}

Our main results, \refThm{finalbd} and \refThm{precisionb}, provide a complete
complexity analysis of the algorithm given in \cite{yap-sagraloff-sharma:cluster:13}. Our intermediate results
show that there is much scope for improvement in the algorithm.

For example, \refLem{term} shows that  scaling in \texttt{firstC} is not required. 
In fact, not scaling  will allow us to terminate with
larger boxes. This  matches with similar results in the polynomial setting \cite{batra-sharma:jsc:17}.
The bound on the subdivision tree shows that the main work done by the algorithm
is in exclusion, and Pellet's test is not a very efficient means to do that. It seems to us that
a winding number argument will be a better means to achieve that, since it measures the
change in $\ln f(z)$ around a closed contour, and when we are  sufficiently far away from a root  we should not 
observe any significant change. The challenge is to develop a rigorous way of doing this without resorting to
$\eps$-cutoffs.

Finally let us mention cluster identification via Smale's
$\alpha$-theory. Zero-inclusion near a cluster center as well as the
distance to the next zero outside the natural cluster may be detected
via comparison of coefficient functions. This may be used to
approximate and locate \emph{groups} of zeros of analytic functions as
in \cite{giusti+2:zeros-analytic:05} and reduce the number of calls
to the exclusion predicate. The approximation in \cite{giusti+2:zeros-analytic:05} is done via a
modified Newton's method (also known as Schr{\"o}der's iteration). The
employed estimates certify that one absolute term in a local function
expansion essentially outweighs other parts of the absolute sum. This
is related to Pellet's test, and it comes as no surprise that methods
combining the latter with Newton's method and its generalizations (as in
\cite{yakoubsohn:bisection-analysis:05} and
\cite{giusti+2:zeros-analytic:05}) have already been proposed. The
combination of these two approaches for polynomial root isolation was
discussed and analyzed recently in \cite{becker+3:cisolate:18}. 

One direction for future work would be a combination of bisection with
non-uniform subdivision driven by a modified Newton's method (similar
to the work on polynomials in \cite{batra-sharma:jsc:17}). This
combination could also be approached in the fashion of
\cite{giusti+2:zeros-analytic:05} by using generalizations of Newton's
method.  

\end{document}